\documentclass[%
superscriptaddress,
 amsmath,amssymb,
 aps,
 prr,
 a4paper,
 twocolumn,
]{revtex4-2}

\usepackage{mathtools}
\usepackage{graphicx}
\usepackage{dcolumn}
\usepackage{bm}
\usepackage[dvipsnames]{xcolor}

\usepackage{csquotes}
\MakeOuterQuote{"}

\usepackage{aliascnt}

\begin{document}

\newcommand{\E}{\mathcal{E}}
\newcommand{\G}{\mathcal{G}}
\newcommand{\Lag}{\mathcal{L}}
\newcommand{\M}{\mathcal{M}}
\newcommand{\N}{\mathcal{N}}
\newcommand{\U}{\mathcal{U}}
\newcommand{\R}{\mathcal{R}}
\newcommand{\F}{\mathcal{F}}
\newcommand{\V}{\mathcal{V}}
\newcommand{\C}{\mathcal{C}}
\newcommand{\I}{\mathcal{I}}
\newcommand{\s}{\sigma}
\newcommand{\up}{\uparrow}
\newcommand{\dw}{\downarrow}
\newcommand{\h}{\hat{{H}}}
\newcommand{\himp}{\hat{h}}
\newcommand{\g}{\mathcal{G}^{-1}_0}
\newcommand{\D}{\mathcal{D}}
\newcommand{\A}{\mathcal{A}}
\newcommand{\projs}{\hat{\mathcal{S}}_d}
\newcommand{\proj}{\hat{\mathcal{P}}_d}
\newcommand{\K}{\textbf{k}}
\newcommand{\Q}{\textbf{q}}
\newcommand{\T}{\tau_{\ast}}
\newcommand{\io}{i\omega_n}
\newcommand{\eps}{\varepsilon}
\newcommand{\+}{\dag}
\newcommand{\su}{\uparrow}
\newcommand{\giu}{\downarrow}
\newcommand{\0}[1]{\textbf{#1}}
\newcommand{\dagga}{{\phantom{\dagger}}}
\newcommand{\ca}{c^{\phantom{\dagger}}}
\newcommand{\cc}{c^\dagger}

\newcommand{\pa}{{p}^{\phantom{\dagger}}}
\newcommand{\pc}{{p}^\dagger}

\newcommand{\fa}{f^{\phantom{\dagger}}}
\newcommand{\fc}{f^\dagger}

\newcommand{\da}{{d}^{\phantom{\dagger}}}
\newcommand{\dc}{{d}^\dagger}

\newcommand{\bF}{\mathbf{F}}
\newcommand{\bD}{\mathbf{D}}

\newcommand{\bR}{\mathbf{R}}
\newcommand{\bQ}{\mathbf{Q}}
\newcommand{\bq}{\mathbf{q}}
\newcommand{\bqp}{\mathbf{q'}}
\newcommand{\bk}{\mathbf{k}}
\newcommand{\bh}{\mathbf{h}}
\newcommand{\bkp}{\mathbf{k'}}
\newcommand{\bp}{\mathbf{p}}
\newcommand{\bL}{\mathbf{L}}
\newcommand{\bRp}{\mathbf{R'}}
\newcommand{\bx}{\mathbf{x}}
\newcommand{\bX}{\mathbf{X}}
\newcommand{\by}{\mathbf{y}}
\newcommand{\bz}{\mathbf{z}}
\newcommand{\br}{\mathbf{r}}
\newcommand{\Ima}{{\Im m}}
\newcommand{\Rea}{{\Re e}}
\newcommand{\Pj}[2]{|#1\rangle\langle #2|}
\newcommand{\ket}[1]{\vert#1\rangle}
\newcommand{\bra}[1]{\langle#1\vert}
\newcommand{\setof}[1]{\left\{#1\right\}}
\newcommand{\fract}[2]{\frac{\displaystyle #1}{\displaystyle #2}}
\newcommand{\Av}[2]{\langle #1|\,#2\,|#1\rangle}
\newcommand{\av}[1]{\langle #1 \rangle}
\newcommand{\Mel}[3]{\langle #1|#2\,|#3\rangle}
\newcommand{\Avs}[1]{\langle \,#1\,\rangle_0}
\newcommand{\eqn}[1]{(\ref{#1})}
\newcommand{\Tr}{\mathrm{Tr}}

\newcommand{\bba}{b^{\phantom{\dagger}}}
\newcommand{\bbc}{b^\dagger}

\title{Active Learning approach to simulations of Strongly Correlated Matter with the Ghost Gutzwiller Approximation}

\author{Marius S. Frank}
\affiliation{
Department of Chemistry, Aarhus University, Langelandsgade 140, 8000 Aarhus C
}

\author{Denis G.\ Artiukhin}
\affiliation{Institut f\"ur Chemie und Biochemie, Freie Universit\"at Berlin, Arnimallee 22, 14195 Berlin, Germany
}

\author{Tsung-Han Lee}
\affiliation{
Department of Physics, National Chung Cheng University, Chiayi 62102, Taiwan
}

\author{Yongxin Yao}
\affiliation{Ames National Laboratory, U.S. Department of Energy, Ames, Iowa 50011, USA}
\affiliation{Department of Physics and Astronomy, Iowa State University, Ames, Iowa 50011, USA}

\author{Kipton Barros}
\affiliation{
Theoretical Division and CNLS, Los Alamos National Laboratory, Los Alamos, New Mexico 87545, USA
}

\author{Ove Christiansen}
\affiliation{
Department of Chemistry, Aarhus University, Langelandsgade 140, 8000 Aarhus C}

\author{Nicola Lanat\`{a}}
\altaffiliation{Corresponding author: nxlsps@rit.edu}
\affiliation{
School of Physics and Astronomy, Rochester Institute of Technology,
84 Lomb Memorial Drive, Rochester, New York 14623, USA}
\affiliation{Center for Computational Quantum Physics, Flatiron Institute, New York, New York 10010, USA}

\date{\today}

\begin{abstract}
Quantum embedding (QE) methods such as the Ghost Gutzwiller Approximation (gGA) offer a powerful approach to simulating strongly-correlated systems, but come with the computational bottleneck of computing the ground state of an auxiliary embedding Hamiltonian (EH) iteratively.
In this work, we introduce an active learning (AL) framework integrated within the gGA to address this challenge. 
The methodology is applied to the single-band Hubbard model and results in a significant reduction in the number of instances where the EH must be solved.
Through a principal component analysis (PCA), we find that the EH parameters form a low-dimensional structure that is largely independent of the geometric specifics of the systems, especially in the strongly-correlated regime.
Our AL strategy enables us to discover this low-dimensionality structure on the fly, while leveraging it for reducing the computational cost of gGA, laying the groundwork for more efficient simulations of complex strongly-correlated materials.
\end{abstract}

\maketitle

\section{Introduction}

At present, most quantitative simulations of quantum matter utilize standard approximations to density functional theory (DFT)~\cite{Hohenberg1964,Kohn1965}. However, these approximations face limitations when simulating the properties of strongly-correlated systems, which are solids and molecules where electrons are localized around specific atomic sites and have intensified interactions due to spatial confinement. This issue is especially relevant in materials containing transition metals from the 3d series and to lanthanides and actinides. 
To address this challenge, various QE~\cite{kent2018toward,Sun2016} many-body techniques have been developed. Methods such as dynamical mean-field theory (DMFT)~\cite{DMFT,dmft_book,Held-review-DMFT,Anisimov_DMFT,LDA+U+DMFT}, density matrix embedding theory (DMET)~\cite{Knizia2012,Wouters2016}, rotationally invariant slave boson theory (RISB)~\cite{Georges-RISB,rotationally-invariant_SB,Lanata2016},
and the multi-orbital Gutzwiller approximation (GA)~\cite{Gutzwiller3,GA-infinite-dim,Fang,Ho,Gmethod,Our-PRX,Bunemann,Attaccalite},
are now widely used for quantitatively simulating strongly-correlated systems. 
Recently an extension of the GA, denoted as the gGA~\cite{Lanata-ghost-2017,gGA-Marius}, has been developed.
The gGA framework incorporates auxiliary Fermionic degrees of freedom to enrich the variational space. Notably, gGA has demonstrated accuracy that is comparable to DMFT~\cite{Lanata-ghost-2017,gGA-Marius,mejutozaera2023efficient,lee2023accuracy,Lee_2022}, indicating that it might serve as an advantageous alternative, especially when aiming for a combination of accuracy and computational manageability.

However, all of the available QE many-body techniques pose a computational burden for emerging applications in materials discovery, where computational efficiency is crucial for reducing both the time and cost of material development.
The main reason lies in their common QE algorithmic structure, that requires the iterative solution of an EH for each correlated fragment in the system, constituting the most computationally intensive step~\cite{NicolaPRX,Lanata-gDMET,Ayral_2017,RISB_DMET_Lee_2019}.
Addressing this bottleneck could enable accurate simulations of strongly-correlated materials at computational costs comparable to traditional approximations to DFT.

In prior work, a machine-learning-based solution to this problem was proposed both in the context of DMFT~\cite{Millis_ML_2014,Weber-ML} and in the context of the GA~\cite{Rogers2021}, exploiting the observation that the mathematical structure of the EH is determined solely by the electron shell structure of the impurity, thus being consistent across diverse materials and molecules. 
This intrinsic commonality that arises when solving the EH across different materials and molecules, which we refer to as
"universality," suggests that machine learning (ML) techniques could, in principle, be trained \emph{once and for all} to solve the EH problem, thereby bypassing the computational bottleneck of all subsequent QE simulations.
In particular, in Ref.~\cite{Rogers2021} a method combining the GA method with a ML algorithm, termed ``n-KRR,'' demonstrated success in implementing this program for a series of actinide systems. However, this achievement was enabled by the possibility to specifically conjecture the physically-relevant range of training data for these materials ---an advantage that is not generally available. Indeed, this represents the primary barrier to overcome for extending this approach to general many-body systems: it is generally impossible to preemptively determine which training data should be generated. Consequently, a different approach is required to make this ML strategy universally applicable.

To overcome the challenge of determining a priori training data, here we introduce an active learning methodology that marries probabilistic ML techniques ---specifically, a recent extension of Gaussian Process Regression (GPR)--- with the gGA framework. 
As new gGA calculations proceed, our active learning model continuously evaluates new instances of the EH problem, adaptively updating and refining its own training set based on the level of uncertainty in its predictions. This strategy eliminates the need for a predefined training set and ensures that only physically relevant data are gathered throughout the computational process.

We benchmark our method using the single-band Hubbard model across varying geometries and interaction strengths, thereby significantly reducing the required number of explicit EH calculations.
Using a PCA, we show that the EH parameters explored throughout these calculations have a low-dimensional structure, largely independent of the specific lattice configurations, particularly in strongly-correlated regimes. We discuss how such inherent low-dimensional structure of the parameter space opens a path for computational techniques commonly found in computer science,
underlining the potential of our active learning strategy to generalize across a wide array of strongly-correlated materials in future work.

\section{Model and $\text{gGA}$ method}

This section aims to lay the foundation for the subsequent development of the QE algorithmic structure. We begin by introducing the single-band Hubbard model, that we employ in our benchmark calculations. We then present the formulation of the gGA. A primary focus of gGA is to iteratively solve for the ground state of an EH, an essential component of the QE approach.

\subsection{The single-band Hubbard model}

For clarity, in this work we present the formalism underlying our AL framework focusing on a generic single-band Hubbard Hamiltonians represented as follows:
\begin{align}
    \hat{H} &=  \sum_{{i,j=1 \atop i \neq j}}^{\N}
    \sum_{\sigma=\uparrow,\downarrow}  t_{ij} 
   \cc_{i\sigma}\ca_{j\sigma} 
   + \sum_{i=1}^{\N}\frac{U}{2}\left(\hat{n}_{i}-1\right)^2 
   -\mu\sum_{i=1}^{\N}\hat{n}_i
   \,,
   \label{Hubbard_model0}
\end{align}
where $\N$ is the number of system fragments, $\cc_{i\sigma}$ and $\ca_{i\sigma}$ are Fermionic creation and annihilation operators, $\sigma$ is a spin label, $i$ and $j$ are fragment labels,  
$\mu$ is the chemical potential,
$U$ is the interaction strength and $\hat{n}_i = \sum_{\sigma} \cc_{i\sigma}\ca_{i\sigma}$ is the number operator for the system fragment $i$, and $t_{ij}$ is a generic hopping matrix, with arbitrary entries.

\subsection{The gGA Lagrange function}

Specializing the theory presented in Refs.~\cite{Lanata-ghost-2017,gGA-Marius} to the single-orbital Hubbard equation given by Eq.~\eqref{Hubbard_model0}, and focusing on solutions preserving both spin and translational symmetries, we find that the ground state in the gGA can be obtained by extremizing the following Lagrange function:
\begin{align}
&\Lag[
{\Phi},E^c;\,  \R,\Lambda;\, \D, \Lambda^{c};\,\Delta, \Psi_0, E;\,\mu]=
\nonumber\\&\;
= \Av{\Psi_0}{\h_{\text{qp}}[\R,\Lambda]}
+E\left(1-\langle\Psi_0|\Psi_0\rangle\right)
\nonumber\\&\;
+\sum_{i=1}^{\mathcal{N}}
\left[\Av{\Phi_i}{\h^i_{\text{emb}}[\D_i,\Lambda_i^c,U,\mu]}
+E_i^c\left(1-\langle \Phi_i | \Phi_i \rangle
\right)\right]
\nonumber\\&\,
-\sum_{i=1}^{\mathcal{N}}\left[
\sum_{\sigma=\uparrow,\downarrow}\sum_{a,b=1}^{B}\big(
[\Lambda_i]_{ab}+[\Lambda^c_i]_{ab}\big)[\Delta_i]_{ab}
\right.
\nonumber\\&\left.
\qquad+\sum_{\sigma=\uparrow,\downarrow}
\sum_{c,a=1}^{B}
\big(
[\D_{i}]_{a} [\R_{i}]_{c}
\left[\Delta_i(\mathbb{I}-\Delta_i)\right]^{\frac{1}{2}}_{ca}
+\text{c.c.}\big)
\right]\,,
\label{Lag-SB-emb}
\end{align}
where $\mathbb{I}$ is the identity matrix, the integer number \(B\) controls the size of the gGA variational space and, in turn, the precision of the gGA approach, \(E\) and \(E_i^c\) are scalars, and \(\Delta_i\), \(\Lambda_i\), and \(\Lambda_i^c\) are \(B\times B\) Hermitian matrices. Additionally, \(\D_i\) and \(\R_i\) are \(B \times 1\) column matrices.
The so-called "quasiparticle Hamiltonian" (\(\h_{\text{qp}}\)) and EH (\(\h^i_{\text{emb}}\)) are defined as:
\begin{align}
    \hat{H}_{\text{qp}}[\R,\Lambda]&=\sum_{i=1}^{\mathcal{N}}\sum_{a,b=1}^{{B}}
    \sum_{\sigma=\uparrow,\downarrow}
    [\Lambda_i]_{ab}\,\fc_{ia\sigma}\fa_{ib\sigma}
    \nonumber\\
    &+\sum_{{i,j=1 \atop i \neq j}}^{N} \sum_{a,b=1}^{B}\sum_{\sigma=\uparrow,\downarrow}[\R_i t_{ij}\R_j^\dagger]_{ab}\,\fc_{ia\sigma}\fa_{jb\sigma}
    \,,
    \label{Hqp}
    \\
    \h^i_{\text{emb}}[\D_i,\Lambda_i^c,U,\mu]&= 
    \frac{U}{2}\left(\hat{n}_i-1\right)^2
    -\mu\, \hat{n}_i
    \nonumber\\
    &+ \sum_{a=1}^{B}\sum_{\sigma=\uparrow,\downarrow}\left[[\D_i]_{a}\,{c}^{\dagger}_{i\sigma}{b}^{\phantom{\dagger}}_{ia\sigma}
    +\text{H.c.}\right]\nonumber\\  &+\sum_{a,b=1}^{B}\sum_{\sigma=\uparrow,\downarrow}[\Lambda^c_i]_{ab}\, {b}^{\phantom{\dagger}}_{ib\sigma}{b}^{\dagger}_{ia\sigma}
   \,.
   \label{hemb}
\end{align}
Here the vector $\ket{\Phi_i}$ is the most general embedding state for the fragment $i$, i.e., the most general state within the Fock space spanned by the $2(B+1)$ modes 
$\cc_{i\sigma}$ and ${b}^{\dagger}_{ia\sigma}$, with $(B+1)$ Fermions in total (half-filled).
The vector $\ket{\Psi_0}$ is the most general single-particle state within the so-called "quasi-particle" space, spanned by the $2B\N$ modes $\fc_{ia\sigma}$.

For \(B=1\), Eq.~\eqref{Lag-SB-emb} reduces to the standard GA Lagrange function.
In this work we set $B=3$, which proved to be sufficient for capturing the ground-state properties with accuracy comparable to DMFT for the ground-state properties~\cite{Lanata-ghost-2017,gGA-Marius,mejutozaera2023efficient,lee2023accuracy,Lee_2022}.

\subsection{Gauge Invariance and Physical Observables}\label{sec:gauge}

It can be readily verified that the gGA Lagrangian is invariant with respect to the following gauge transformation:
\begin{align}
    \ket{\Psi_0} &\rightarrow \U^{\dagger}\left(\theta\right)\ket{\Psi_0} 
    \label{g1}
    \\
    \ket{\Phi_i} &\rightarrow U_i^{\dagger}\left(\theta\right)\ket{\Phi_i} \\
    \R_i &\rightarrow u_i^{\dagger}\left(\theta\right)\R_i \\
    \D_i &\rightarrow  u_i^\mathrm{T}\left(\theta\right)\D_i \\
    \Delta_i &\rightarrow  u_i^\mathrm{T}\left(\theta\right)\Delta_i  \, u_i^*\left(\theta\right) \\
    \Lambda_i &\rightarrow u_i^{\dagger}\left(\theta\right)\Lambda_i \, u_i \left(\theta\right) \\
    \Lambda_i^c &\rightarrow u_i^{\dagger}\left(\theta\right) \Lambda_i^c\, u_i\left(\theta\right)
    \label{g7}
    \,,
\end{align}
with:
\begin{align}
    u_i\left(\theta_i\right) &= e^{i\theta_i} \\
    U_i\left(\theta_i\right) &= e^{i\sum_{a,b=1}^{B}[\theta_i]_{ab} {b}^{\dagger}_{ia}{b}^{\phantom{\dagger}}_{ib}} \\
    \U\left(\theta\right) &= e^{i\sum_{i=1}^\N\sum_{a,b=1}^{B}[\theta_i]_{ab} \fc_{ia}\fa_{ib}} \,,
\end{align}
where $\theta=(\theta_1,..,\theta_\N)$, $\theta_i$ are $B\times B$ Hermitian matrices and the superscript "$\mathrm{T}$" denotes the transpose, while the superscript "$*$" denotes the complex conjugate.
The name "gauge" here refers to the fact that modifications of the parameters generated by such a gauge transformation do not influence any physical observable, which can be extracted from the variational parameters that extremize the Lagrange function.

For completeness, below we write explicitly how the physical observables that we calculate in this paper
are computed as a function of the variational parameters, based on
the theoretical framework derived in previous work~\cite{Lanata-ghost-2017,gGA-Marius}.

The total energy of the system is given by the Lagrange function value after extremization (which is gauge invariant).
The expectation values for local observables are encoded in \(\ket{\Phi_i}\). In particular, the local double-occupancy expectation value in the gGA ground state is given by the following gauge-invariant expression:
\[
\langle \hat{n}_{i\uparrow}\hat{n}_{i\downarrow} \rangle_{\text{gGA}} = 
\Av{\Phi_i}{\hat{n}_{i\uparrow}\hat{n}_{i\downarrow}}\,,
\]
where \(\hat{n}_{i\sigma} = \cc_{i\sigma}\ca_{i\sigma}\).
To calculate the quasi-particle weight, it is convenient to express the variational parameters in a gauge where $[\Lambda_i]_{ab}=[l_i]_a\delta_{ab}$ (which always exists, since $\Lambda_i$ is Hermitian).
In this gauge, the mathematical expression for the quasi-particle weight is the following:
\begin{align}
    Z_i &= \left[ 1 - \frac{\partial \Sigma_i}{\partial\omega}\right]^{-1}_{\omega= 0}
    \\
    &= \frac{\left([l_i]_2 \, [l_i]_3 \, [\R_i]_1^2 + [l_i]_1 \, [l_i]_3 \, [\R_i]_2^2 + [l_i]_1 \, [l_i]_2 \, [\R_i]_3^2\right)^2}
    {[l_i]_1^2 \, [l_i]_3^2 \, [\R_i]_2^2 + [l_i]_2^2 \left([l_i]_3^2 \, [\R_i]_1^2 + l_1^2 \, [\R_i]_3^2\right)}  \nonumber
    \,.
\end{align}
where $\Sigma_i(\omega)$ is the self-energy.

\section{Formulation of the ML problem}

A pivotal insight at the base of the ML approach proposed in this work is that the problem of extremizing the Lagrange function in Eq.~\eqref{Lag-SB-emb} can be formally tackled by 
first solving for $\ket{\Phi_i}$ and $E^c_i$ as a function of 
the other parameters.
This amounts to replacing the original variational problem of extemizing $\Lag$ in Eq.~\eqref{Lag-SB-emb} with the problem of extremizing the following Lagrange function: 
\begin{align}
&\bar{\Lag}[
\R,\Lambda;\, \D, \Lambda^{c};\,\Delta, \Psi_0, E;\,\mu]=
\label{Lag-SB-emb-ML}
\\&\;
= \Av{\Psi_0}{\h_{\text{qp}}[\R,\Lambda]}
+E\left(1-\langle\Psi_0|\Psi_0\rangle\right)
\nonumber\\&\;
+\sum_{i=1}^{\mathcal{N}}
\bar{E}^c(\D_i,\Lambda_i^c,U,\mu)
\nonumber\\&\,
-\sum_{i=1}^{\mathcal{N}}\left[
\sum_{\sigma=\uparrow,\downarrow}\sum_{a,b=1}^{B}\big(
[\Lambda_i]_{ab}+[\Lambda^c_i]_{ab}\big)[\Delta_i]_{ab}
\right.
\nonumber\\&\left.
\qquad+\sum_{\sigma=\uparrow,\downarrow}
\sum_{c,a=1}^{B}
\big(
[\D_{i}]_{a} [\R_{i}]_{c}
\left[\Delta_i(\mathbb{I}-\Delta_i)\right]^{\frac{1}{2}}_{ca}
+\text{c.c.}\big)
\right]
\nonumber
\,,
\end{align}
where:
\begin{align}
&\bar{E}^c(\D, \Lambda^c, U,\mu) =  
\left\langle{\h^i_{\text{emb}}[\D,\Lambda^c, U,\mu]}\right\rangle_{\D,\Lambda^c, U,\mu}
\,,
\label{eq:Ec}
\end{align}
$\h^i_{\text{emb}}$ is the EH defined
in Eq.~\eqref{hemb}, and the expectation value is taken with respect to the corresponding half-filled ground state $\ket{\Phi_i}$.
Note that the subscript $i$
is not present in the function $\bar{E}^c$
of Eq.~\eqref{Lag-SB-emb-ML}, highlighting the fact that the function does \emph{not} depend on it.

A key property of the function \( \bar{E}^c(\D, \Lambda^c, U,\mu) \) is that it is "universal," in the sense that its definition is irrespective of the details of the model system under study. For example, when applying gGA to any of the Hubbard models defined in Eq.~\eqref{Hubbard_model}, the form of \( \bar{E}^c(\D, \Lambda^c, U,\mu) \) would remain consistent, independent of the specific lattice structure or the numerical values of the hopping matrix \( t \).
Therefore, if one could learn the energy function \( \bar{E}^c(\D, \Lambda^c, U,\mu) \), along with its gradient, the computational cost of extremizing $\bar{\Lag}$ would be significantly alleviated.
The purpose of this work is to derive a ML model to accomplish this goal.

\subsection{The EH Universal Function}\label{sec:Ec}

The Lagrange equations obtained by extremizing the Lagrange function in Eq.~\eqref{Lag-SB-emb-ML} can be solved iteratively~\cite{Lanata-ghost-2017,gGA-Marius}. The specific algorithm employed in this work is detailed in the Supplemental Material~\cite{supplemental_material}.

For the purposes of this paper, the essential point is that the algorithmic structure involves iteratively 
evaluating \( \bar{E}^c(\D_i, \Lambda_i^c, U,\mu) \) and its gradient, as the parameters \( \D_i \) and \( \Lambda_i^c \) are updated at each step. 
Specifically, the computational bottleneck lies in the evaluation of the ground-state single-particle density matrix elements of each fragment $i$:
\begin{align}
\frac{\partial \bar{E}^c}{\partial [\D_i]_a} &= 2[\rho^{\text{hyb}}_i]_{a}
= 2\left\langle{\sum_\sigma c^\dagger_{i\sigma} b_{ia\sigma}}\right\rangle_{\D_i,\Lambda_i^c, U,\mu}
 \\
\frac{\partial \bar{E}^c}{\partial [\Lambda^c_i]_{ab}} &= [\rho^{\text{bath}}_i]_{ab}
= \left\langle{\sum_\sigma b_{ib\sigma} b^\dagger_{ia\sigma} }\right\rangle_{\D_i,\Lambda^c_i, U,\mu}
\,,
\end{align}
where the identities above hold true because of the Hellmann--Feynman theorem.

\subsection{Reducing the complexity of the learning problem}
\label{sec:red-comp}

In the previous subsection we have introduced the function \( \bar{E}^c(\D, \Lambda^c, U,\mu) \). Note that, since $\D$ and $\Lambda^c$ are generally complex and $\Lambda^c$ is Hermitian,
$\bar{E}^c$ is a function of $2B+B^2+2$ real parameters.

In this section we show that it is possible to reduce the problem of learning 
the aforementioned universal EH energy function to the problem of learning 
the following function of only $2B$ real variables:
\begin{align}
    &\E(\tilde{\D}_1,..,\tilde{\D}_B,\tilde{\Lambda}^c_{11},..,\tilde{\Lambda}^c_{BB})=
    \nonumber\\&\quad
    \left\langle
    \h_{\text{emb}}[\tilde{\D}_1,..,\tilde{\D}_B,\tilde{\Lambda}^c_{11},..,\tilde{\Lambda}^c_{BB}]
    \right\rangle_{\tilde{\D}_1,..,\tilde{\D}_B,\tilde{\Lambda}^c_{11},..,\tilde{\Lambda}^c_{BB}}
    \label{def:mathcalEc}
    \,,
\end{align}
representing the ground-state energy of the following Hamiltonian:
\begin{align}
    &\h_{\text{emb}}[\tilde{\D}_1,..,\tilde{\D}_B,\tilde{\Lambda}^c_{11},..,\tilde{\Lambda}^c_{BB}]
    \nonumber\\
    &\qquad= \frac{1}{2}\left(\hat{n}-1\right)^2
    \nonumber\\
    &\qquad+ \sum_{a=1}^{B}\sum_{\sigma=\uparrow,\downarrow}\left[\tilde{\D}_{a}\,{c}^{\dagger}_{\sigma}{b}^{\phantom{\dagger}}_{a\sigma}
    +\text{H.c.}\right]\nonumber\\  &\qquad+\sum_{a=1}^{B}\sum_{\sigma=\uparrow,\downarrow}\tilde{\Lambda}^c_{aa}\, {b}^{\phantom{\dagger}}_{a\sigma}{b}^{\dagger}_{a\sigma}
    \label{def:Hembtilde}
\end{align}
at half filling.
Furthermore, it is sufficient to learn the function $\mathcal{E}$, defined by Eq.~\eqref{def:mathcalEc}, over the restricted domain such that
$\tilde{\D}_1\geq \tilde{\D}_2\geq ... \tilde{\D}_B$.

This simplification is made possible by 
the fact that the EH function $\bar{E}^c$, defined in Sec.~\ref{sec:Ec},
satisfies the following general properties:
\begin{itemize}
    \item Invariance of half-filled ground state $\ket{\Phi}$
    under simultaneous shift of impurity and bath energies:
    \begin{equation}
        \bar{E}^c(\D, \Lambda^c, U,\mu)=\bar{E}^c(\D, \Lambda^c+\mu \mathbb{I}, U,0)-\mu
        \,;
    \end{equation}
    \item Linear homogeinity: 
    \begin{equation}
        \bar{E}^c(x\D, x\Lambda^c, xU,x\mu)=x\bar{E}^c(\D, \Lambda^c, U,\mu)\quad\forall\, x\,;
        \label{eq:LinHom}
    \end{equation}
    \item Gauge invariance:
    \begin{equation}
        \bar{E}^c(u^\mathrm{T}\D, u^\dagger\Lambda^c u, U,\mu)=\bar{E}^c(\D, \Lambda^c, U,\mu)\quad  \forall\, u\,;
        \label{eq:GaugeInv}
    \end{equation}
\end{itemize}
where $x$ is any real number and $u$ is any $B\times B$ unitary matrix (i.e., a gauge transformation, see Sec.~\ref{sec:gauge}).
From the properties above it follows that:
\begin{align}
    \bar{E}^c(\D, \Lambda^c, U,\mu)&=
    \bar{E}^c(\D, \Lambda^c+\mu \mathbb{I}, U,0)+\mu
    \label{def:tildeEc}
    \\ 
    &=U \bar{E}^c(\tilde{\D},\tilde{\Lambda}^c,1,0)+\mu
    \nonumber\\ 
    &=U\mathcal{E}(\tilde{\D}_1,..,\tilde{\D}_B,\tilde{\Lambda}^c_{11},..,\tilde{\Lambda}^c_{BB})+\mu
    \nonumber
    \,,
\end{align}
where:
\begin{align}
    \label{lcparam}
    \tilde{\Lambda}^c &= \frac{1}{U} u^{\dagger}(\Lambda^c+\mu \mathbb{I}) u
    \\
    \label{Dparam}
    \tilde{\D} &= \frac{1}{U}  u^\mathrm{T} \D \,,
\end{align}
and the unitary matrix $u$ is a gauge transformation chosen in such a way that $\tilde{\Lambda}^c$ is diagonal and the entries of $\tilde{\D}$ are real and sorted in descending order,
as detailed in the Supplemental Material~\cite{supplemental_material}.

Another important consequence of Eqs.~\eqref{eq:LinHom} and \eqref{eq:GaugeInv} is that, as shown in the Supplemental Material~\cite{supplemental_material}, also the gradient of $\bar{E}^c$ is fully encoded in the gradient of $\mathcal{E}$, which is given by the following equations:
\begin{align}
\frac{\partial \E}{\partial \tilde{\D}_a} &= 2\tilde{\rho}^{\text{hyb}}_{a}=2
\left\langle{\sum_\sigma c^\dagger_\sigma b_{a\sigma}}\right\rangle_{\tilde{\D}_1,..,\tilde{\D}_B,\tilde{\Lambda}^c_{11},..,\tilde{\Lambda}^c_{BB}}
\label{eq:rhohyb2}
\\
\frac{\partial \E}{\partial \tilde{\Lambda}^c_{aa}} &= \tilde{\rho}^{\text{bath}}_{aa}
= \left\langle{\sum_\sigma b_{a\sigma} b^\dagger_{a\sigma} }\right\rangle_{\tilde{\D}_1,..,\tilde{\D}_B,\tilde{\Lambda}^c_{11},..,\tilde{\Lambda}^c_{BB}}
\label{eq:rhobath2}
\,,
\end{align}
where the expectation values are taken with respect to the ground state 
of the EH defined in Eq.~\eqref{def:Hembtilde}.

\subsection{Summary}\label{sec:summary-defml}

In summary, in this section we have reduced the problem of solving the gGA equations to an iterative procedure which consists of evaluating iteratively the functions:
\begin{align}
\E(\bX)&=\E(\tilde{\D}_1,..,\tilde{\D}_B,\tilde{\Lambda}^c_{11},..,\tilde{\Lambda}^c_{BB})
    \label{E}
    \\
\bF(\bX)&=\nabla\E(\bX)
    \label{F}
    \,,
\end{align}
where we have introduced the $2B$-dimensional real vector:
\begin{equation}
\bX=(\tilde{\D}_1,..,\tilde{\D}_B,\tilde{\Lambda}^c_{11},..,\tilde{\Lambda}^c_{BB})\,.
\end{equation}
Each evaluation of $\E(\bX)$ and $\bF(\bX)$ requires to calculate the ground state of the EH defined in Eq.~\eqref{def:Hembtilde}, whose dimension is $2^{2(B+1)}$.
This can quickly become the computational bottleneck of the gGA framework as $B$ increases, which we aim to mitigate using ML.

An important point to highlight is the simplification introduced by reducing the problem to the learning of a single scalar function \( \E(\bX) \), which we achieved through the use of the Hellmann--Feynman theorem. 
This approach streamlines computation and reinforces predictive accuracy compared to the method of individually learning each entry of the ground-state single-particle density matrix of the EH (which is the technique we previously employed in Ref.~\cite{Rogers2021}).
First, learning a single scalar function is computationally less demanding than learning an array of functions corresponding to each matrix element. 
Second, this method also automatically enforces inherent prior information about these specific functions, such as the condition that: 
\begin{equation}
\nabla \times \bF(\bX) = 0
\,,
\end{equation}
i.e., \( \bF(\bX) \) is conservative, thereby enhancing the overall predictive accuracy of the model.

Interestingly, from the mathematical perspective, the problem outlined above bears a striking resemblance
to a specific successful ML application, namely learning force fields for accelerating molecular dynamics
simulations~\cite{Gabi-2017,Xin-2017,Kipton-2019}. In fact, in both instances, the challenge revolves around learning the total energy $\E(\bX)$ of a system and its
gradient $\bF(\bX)$ from computational data. 
On the other hand, there are 2 key differences:
\begin{itemize}
    \item \emph{Universality}: Within our framework, if we could learn once and for all 
    $\E(\bX)$, the resulting model could be used to bypass such computational bottleneck for \emph{any} system involving single-orbital fragments (e.g., all models
    of the form represented in Eq.~\eqref{Hubbard_model}, for all  hopping matrices $t_{ij}$ and for any values of $U$ and $\mu$).
    
    \item \emph{Domain structure}: In gGA the sequence of points $\bX_\alpha$ explored throughout the computation of any given model always converges towards the specific $\bar{\bX}$ realizing the corresponding solution. Therefore, the majority of these points, where $\E(\bx)$ needs to be learned,
    gravitate around $\bar{\bX}$, instead of being spread around the whole parameters space.
\end{itemize}

These unique characteristics suggest a more tractable learning problem. Specifically, we are not compelled to learn the universal function \( \E(\bX) \) over its entire domain, hereafter referred to as the \emph{"ambient space"}. 
Rather, we can confine our attention to a subset of parameters situated in the proximity of the ground states of physical models. These ground states encapsulate the possible physical embeddings, i.e., the feasible interactions that fragments can have with their environment in the ground state of physical systems. 
Such parameters presumably constitute a limited fraction of the ambient space. 

Therefore, we are confronted with a dual challenge. The first is learning the function \( \E(\bX) \) over this restricted domain, hereafter referred to as the \emph{"latent space."} The second is concurrently unveiling the structure of this latent space, which is expected to have lower dimensionality than the ambient space. 
The successful completion of the latter task would considerably streamline the overall learning problem, specifically by potentially mitigating the effects of the so-called "curse of dimensionality," where computational cost grows exponentially with the number of dimensions.

In the next section we describe an AL framework for learning $\E(\bX)$ and $\bF(\bX)$, which is specifically tailored for this purpose, capitalizing on the observations above.

\section{Active Learning Framework}

In this section, we outline an AL strategy based on probabilistic ML to overcome the computational challenges delineated in the previous section.

A critical aspect of our strategy is the use of a probabilistic ML model, i.e., a model capable of combining our prior knowledge and observed data to make predictions for \(\E\) and its gradient \(\bF\) as expectation values \( \langle \E(\bX) \rangle \) and \( \langle \bF(\bX) \rangle \) with respect to a suitable probability distribution, and to quantify their uncertainties through standard deviations:
\begin{align}
    \Sigma_0(\bX)  &= \big[\langle \E^2(\bX)\rangle-\langle \E(\bX)\rangle^2\big]^{\frac{1}{2}}
    \\
    \Sigma_i(\bX)  &= \big[\langle F_i^2(\bX)\rangle-\langle F_i(\bX)\rangle^2\big]^{\frac{1}{2}}\,,
\end{align}
 where \( F_i(\bX) \) are the components of \( \bF(\bX) \), with \( i=1,\ldots,2B \), where $2B$ is the dimension of $\bX$.
Furthermore, we require a model capable of learning both the energy function \(\E\) and its gradient \(\bF\) simultaneously, ensuring exactly the consistency between these two quantities.
We achieve this by a recent generalization of GPR, satisfying both of these requirements~\cite{schmitz2020,arti2023}, described in Sec.~\ref{sec:gpr-main} and in the Supplemental Material~\cite{supplemental_material}.

\begin{figure}
    \centering
    \includegraphics[width=0.47\textwidth]{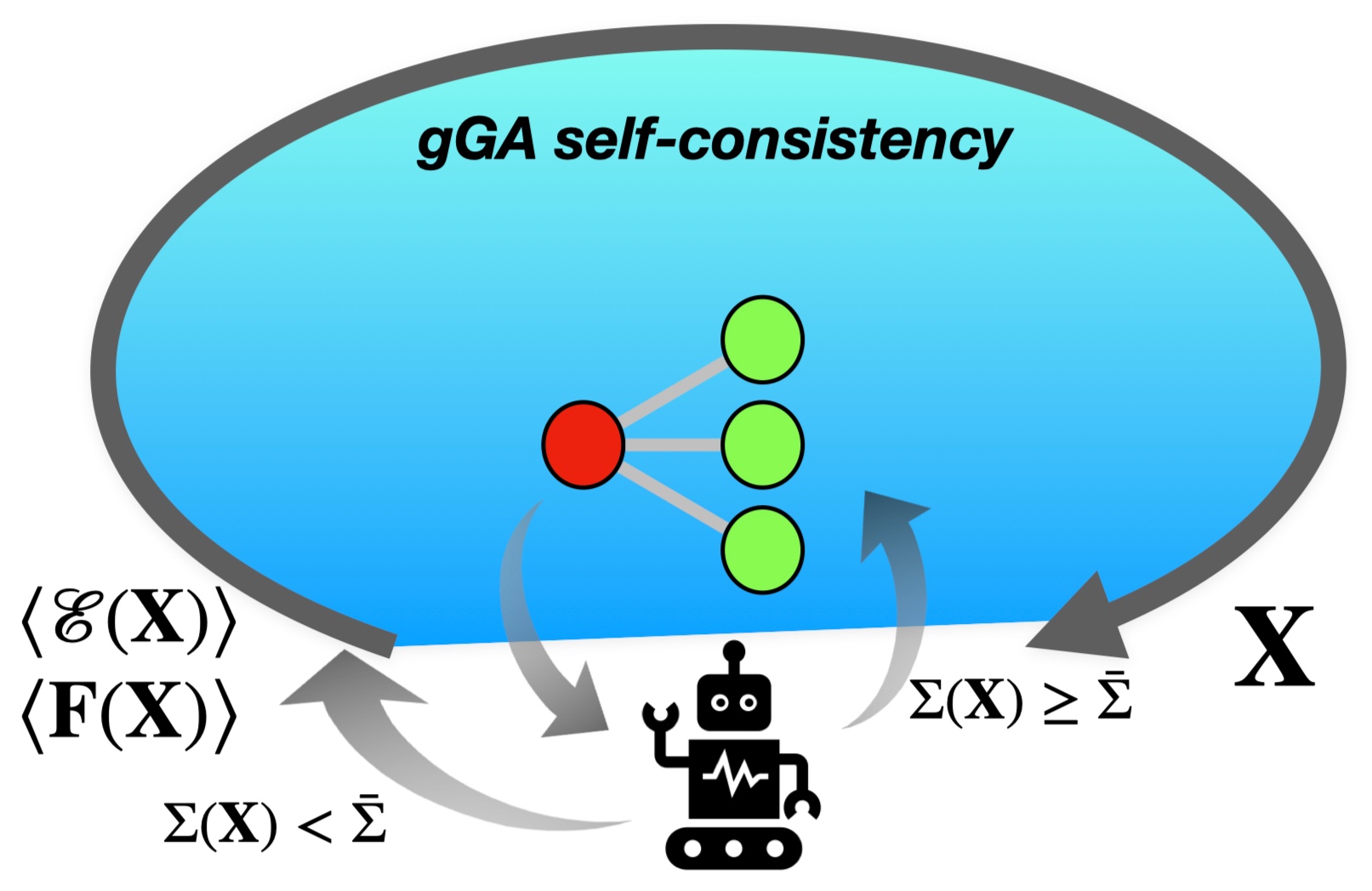}
    \caption{Schematic representation of the AL structure. A machine is trained on the fly while performing gGA calculations. If the uncertainty estimate $\Sigma$ for ML predictions at a given point $\bX$ is within a threshold, the prediction is accepted. Otherwise, the machine database is updated with energy $\mathcal{E}$ and gradient $\mathbf{F}$ data, and a new machine evaluation is performed.}
    \label{Figure1}
\end{figure}

\subsection{gGA+AL Algorithmic Structure}\label{sec:AL-structure}

Given a probabilistic ML method with the requirements listed above in place, we proceed to outline the AL strategy as follows:
\begin{enumerate}
    \item For each self-consistency cycle, the gGA algorithm produces a parameter vector \( \bX \) for every impurity in the system.
    
    \item The probabilistic ML framework predicts the expectation values \( \langle \E(\bX) \rangle \) and \( \langle \bF(\bX) \rangle \). It also estimates the uncertainty quantification of these predictions, termed $\Sigma_0(\bX)$ and $\Sigma_i(\bX)$, respectively.
 
    \item If the existing data can produce a sufficiently accurate prediction for  \( \langle \E(\bX) \rangle \) and \( \langle \bF(\bX) \rangle \) (with respect to pre-established accuracy thresholds), the ML model's outputs are directly employed, thereby bypassing the need for ground state computation of the EH.
    
    \item If the data are insufficient for an accurate prediction corresponding to the EH at the given \( \bX \), new training data are computed for suitable parameters \( \bX_\alpha \). These data points are then added to the database for future use.
    
    \item The output, obtained through either step 3 or 4, is fed back into the QE algorithm to initiate the next iteration. The process is repeated until convergence is achieved.
\end{enumerate}

The salient features of this framework, as depicted in Fig.~\ref{Figure1}, are twofold: 
(1) Since all data are generated through actual gGA calculations, they are inherently situated in the proximity of the "latent space" of physically meaningful embeddings. This circumvents the issue of requiring prior knowledge of the QE latent space, as discussed in Sec.~\ref{sec:summary-defml}.
(2) Owing to the universality of the EH, data collected during any calculation for strongly correlated systems are retained permanently. Consequently, the ML component of our framework becomes increasingly efficient with each new calculation, potentially enhancing the computational efficiency of all subsequent gGA simulations.

The details about the algorithm implementation outlined above are described in the subsections below.

\subsection{Generalized Gaussian Process Regression} \label{sec:gpr-main}

Above in this section we introduced the need for a probabilistic ML model that can learn and predict both the energy function \(\E\) and its gradient \(\bF\), as well as quantify their uncertainties. To meet these requirements, we employ a generalized form of Gaussian Process Regression (GPR)~\cite{schmitz2020,arti2023}, as implemented in a development version of the program package \textit{MidasCpp}~\cite{midascpp}.
The method constructs a so-called "posterior probability distribution" for the function to learn, which is based on a prior probability distribution and available data. 
\begin{itemize}
\item The prior encodes our expectations about the general properties of the function we aim to learn, such as its range and smoothness. 
\item The observed data in our case consist of a database:
\begin{equation}
    \bD=\{\bX_\alpha, \E_\alpha, \bF_\alpha, \sigma_{0\alpha},\boldsymbol{\sigma}_{\alpha}\}
    \label{eq:data}
    \,,
\end{equation}
where \(\bX_\alpha\) are points with evaluated energies \(\E_\alpha\) and gradients \(\bF_\alpha\), \(\sigma_{0\alpha}\) measures the uncertainty associated with the energy data $\E_\alpha$, and $\boldsymbol{\sigma}_{\alpha}=(\sigma_{1\alpha},..,\sigma_{d\alpha})$, where $\sigma_{i\alpha}$ is the uncertainty associated with the $i$-th component of the gradient data $\bF_{i\alpha}$ and $d$ is the dimension of $\bX_\alpha$. 
\end{itemize}

In GPR the prior probability distribution is assumed to be a zero-mean Gaussian. Consequently, it is fully characterized by the so-called "kernel function," which is essentially the correlation function \(\langle \E(\bX) \E(\bX') \rangle_{\text{prior}}\), where the expectation value is taken with respect to the prior probability distribution.
The kernel function we employ in this work is the "square exponential kernel," given by:
\begin{align}
    k(\mathbf{X}, \mathbf{X}') &= \langle \E(\bX) \E(\bX') \rangle_{\text{prior}}
    \nonumber\\
    &=\sigma_f^2 \exp \left( - \frac{(\bX - \bX' )^2}{2l^2} \right)
    \,.
\end{align}
This kernel is governed by two hyperparameters: \(l\) and \(\sigma_f\). The length scale \(l\) is essentially a correlation length, defining the expected "minimum wavelength" or smoothness of the function \(\E(\bX)\). The parameter \(\sigma_f\) specifies the expected amplitude or range of the function, serving as an infrared cutoff. Specifically, \(k(\bX, \bX) = \sigma_f^2\) corresponds to the expected variance \(\langle \E(\bX)^2 \rangle_{\text{prior}}\) of the function.

The posterior probability distribution thus integrates both our prior knowledge and the available observed data, generating predictions for \(\E\) and \(\bF\) that align with the observed data, while utilizing the prior for making predictions elsewhere. In particular, our AL framework requires to compute the following quantities at any given point $\bX$:
\begin{align}
    \bar{\E}(\bX) &= \langle \E(\bX)\rangle_{l,\sigma_f,\bD}
    \\
    \bar{F_i}(\bX) &= \langle F_i(\bX)\rangle_{l,\sigma_f,\bD}
    \\
    \Sigma_0(\bX)  &= \big[\langle \E^2(\bX)\rangle_{l,\sigma_f,\bD}-\langle \E(\bX)\rangle_{l,\sigma_f,\bD}^2\big]^{\frac{1}{2}}
    \\
    \Sigma_i(\bX)  &= \big[\langle F_i^2(\bX)\rangle_{l,\sigma_f,\bD}-\langle F_i(\bX)\rangle_{l,\sigma_f,\bD}^2\big]^{\frac{1}{2}}\,,
\end{align}
where the expectation values are taken with respect to the posterior 
distribution, that depends on the hyperparameters $l,\sigma_f$, as well as the available data $\bD$.
Explicit expressions for these distributions are provided in the Supplemental Material~\cite{supplemental_material}.

It is important to note that, as opposed to standard GPR and the KRR-based method previously used in Ref.~\cite{Rogers2021}, the generalized GPR framework outlined above enforces by construction the condition:
\begin{equation}
    \bar{\bF}(\bX)=\nabla\bar{\E}(\bX)
    \,,
\end{equation}
where \( \bar{F}_i(\bX) \) are the components of \( \bar{\bF}(\bX) \), with \( i=1,\ldots,d \), where \(d = 2B\) is the dimension of $\bX$.
Enforcing exactly these conditions yields more accurate predictions.
On the other hand, as explained in the Supplemental Material~\cite{supplemental_material}, such gain comes with additional computational cost.
Specifically, if the database \( \bD \) contains \( N \)
training data points, making predictions requires to invert a matrix (the so-called "covariance matrix"),
whose size is \( N(d+1)\times N(d+1) \),
while it is only \( N\times N \) in standard GPR.
In light of this, the computational complexity and matrix size present two challenges that need to be carefully addressed:
\begin{itemize}
    \item \textbf{RAM Storage:} The large covariance matrix, of size \( N(d+1)\times N(d+1) \), necessitates considerable memory storage. This can become a significant issue as the number of data loaded in the GPR framework grows.
    \item \textbf{Scalability:} The matrix inversion operation itself has a time complexity of \( \mathcal{O}\left( (N(d+1))^3 \right) \). As \( N \) increases, it becomes computationally burdensome. Specific measures must be incorporated into our active learning framework to ensure its scalability for large databases.
    \item \textbf{Numerical Stability:} The large size of the covariance matrix and closely spaced data points in \( \bD \) can make the matrix inversion prone to numerical issues. Specifically, the matrix can become ill-conditioned, having a large condition number, which is the ratio of the largest to the smallest eigenvalue. This makes the matrix sensitive to small changes, potentially leading to a loss of numerical precision.

\end{itemize}

In the following subsection, we detail how these challenges are addressed in our active learning framework, and also elaborate on our procedure for choosing the hyperparameters \(l\) and \(\sigma_f\).

\subsection{Addressing computational and numerical challenges in AL framework}\label{sec:details}

To cope with the computational and numerical challenges presented by the large covariance matrix, we have designed the following strategy within our AL framework.

Our approach is based on the construction of a hyper cubic lattice discretization of the \(d\)-dimensional parameter space \(\bX\), yielding a set of grid points \(\bX_{i_1,\ldots,i_d}\). Each \(i_\alpha\) index covers all integers, effectively tiling the whole parameter space. We use a lattice spacing \(a=0.0125\), which is set to be much smaller than $1$, taking into account the dimensionless nature of our parameters.
When predicted values for a given \(\bX\) are being requested during the self-consistent embedding computations we find the nearest point on the lattice to this point. 
In addition we also consider this lattice points \(2d\) nearest neighbors on the lattice (corresponding to $i_d \pm 1$ for all $d$). Should any of these $2d+1$ points not already be in the database \(\bD\), they are calculated and added. Note that these calculations are independent of each other and, therefore, can be executed in parallel.
For each evaluation at \(\bX\), the GPR prediction and its associated uncertainties are calculated based solely on these \(2d+1\) data points. We locate these nearest points using a k-d tree-based algorithm to maintain computational efficiency.
This step effectively imposes a "budget" on the number of training data we use for making GPR predictions, thereby controlling the size of the covariance matrix and the associated computational complexity.

Next, we address the setting of the hyperparameter \(\sigma_f\) within this localized framework. For the GPR prediction of each point \(\bX\), the hyperparameter \(\sigma_f\) for the squared exponential kernel is set based on the data as follows:
\begin{align}
    \sigma_f^2 = \frac{1}{2d+1}\sum_{\alpha=1}^{2d+1} \mathcal{E}^2(\bX_\alpha)
    \,,
\end{align}
which is consistent with its interpretation in terms of the prior probability distribution: $\sigma^2_f=\langle\mathcal{E}^2(\bX)\rangle_{\mathrm{prior}}$.
To determine the appropriate value of the hyperparameter \(l\), we employ the following iterative approach:
\begin{enumerate}
    \item Initialize \(l\) at \(l_{\text{init}} = 0.5\). The range for \(l\) is predetermined to be between \(l_{\text{init}}\) and \(l_{\text{final}} = 2.0\), with increments of \(\Delta l = 0.1\).
    
    \item For the current \(l\), calculate \(\Sigma_{\text{max}}\), which is the maximum of the uncertainties \(\Sigma_i(\bX)\) associated with all components of the gradient \(\bF(\bX)\) for \(i=1,..,d\).
    
    \item If \(\Sigma_{\text{max}} < \bar{\Sigma}=10^{-3}\), accept the current GPR prediction for that \(l\) and terminate the loop.
    
    \item Otherwise, increment \(l\) by \(\Delta l\) and return to step 2. If \(l\) exceeds \(l_{\text{final}}\), revert to exact calculations for that specific test point and terminate the loop.
\end{enumerate}
The uncertainty parameters $\sigma_{i\alpha}$, see Eq.~\eqref{eq:data}, have been all set to $10^{-5}$ in all of our calculations.

The rationale underlying this algorithm is to initiate with the smallest \(l\) value, thereby making the least assumptions about the landscape's smoothness and relying more heavily on the data for our predictions. If this \(l\) proves insufficient, we then increment \(l\) in a stepwise manner, each time reassessing the prediction quality. This iterative fine-tuning of \(l\) is in essence a method for optimizing the trade-off between bias and variance, a standard criterion in ML, which ensures that our model is neither too simplistic (high bias; large \(l\)) nor too sensitive to fluctuations in the data (high variance; small \(l\)).
Note also that our choice of \(l\) range is in line with the fact that our parameters \(\bX\) are dimensionless, from which we expect the optimal \(l\) to be around 1. Furthermore, it is consistent with the assumption \(l \gg a\), acknowledging that we cannot resolve scales smaller than the lattice spacing \(a\).
Finally, note that our choice of $\bar{\Sigma}=10^{-3}$ is consistent with the assumption $\bar{\Sigma}\gg\sigma_{i\alpha}$,
acknowledging that it is impossible to make predictions with higher accuracy than the available data.
Thus, the framework above effectively manages challenges related to RAM storage, computational scalability, and numerical stability, while maintaining a balance between computational cost and prediction accuracy.

\section{Results} 

In this section we document the results obtained by applying the AL algorithm described above to gGA calculations on the single-band Hubbard model at half filling. Here, we consider different values for the Hubbard interaction strength \(U\) and hopping parameters \(t_{ij}\) (corresponding to multiple lattice geometries).

We carry out our calculations within the gGA framework set at \(B=3\), which, as proven in previous work, is sufficient for achieving accuracy comparable to DMFT for ground-state properties. As clarified in Sec.~\ref{sec:summary-defml}, in this setting the dimension of the "ambient space" of parameters \(\bX\), where the energy function \(\E(\bX)\) is defined, is \(2B=6\). All training-data evaluations for \(\E(\bX)\) and \(\bF(\bX)\) are calculated using the exact diagonalization (ED) method.

In Subsections~\ref{sec:benchmarkB} and \ref{sec:benchmarkC} we document the efficiency and accuracy obtained in these calculations using our AL approach. In Subsection~\ref{sec:PCA}, we demonstrate that the parameters explored span a low-dimensional latent space and discuss the practical implications of these results, as well as their physical interpretation in relation to Mott physics.
Additional calculations and analysis for the Hubbard model away from half filling are presented in the Supplemental Material~\cite{supplemental_material}.

\subsection{Goal of Benchmark Calculations}\label{sec:benchmarks-goal}

Our aim is to test the AL method within all interaction regimes of the half-filled Hubbard model at zero temperature, including the so-called coexistence region, which is an interval of parameters $U$ featuring a metastable Mott state.
To capture all of these regimes, all calculations are organized into series constructed as follows. Each series starts from a large value of interaction strength \(U_{\text{max}}\) and decreases it in intervals of \(\Delta U\) to a small value \(U_{\text{min}}\). Then, the interaction strength is increased back to \(U_{\text{max}}\) with the same spacing. From now on, we refer to such a series of calculations as a "sweep."

A critical metric for quantifying the efficiency of our gGA+AL approach is the ratio of the number of times new data must be acquired and added to the database during a given calculation, \(N_{\text{data}}\), to the total number of gGA iterations necessary to perform the same calculation without ML, \(N_{\text{iterations}}\):
\begin{equation}
S = \frac{N_{\text{data}}}{N_{\text{iterations}}}\,,
\end{equation}
which we would like to be as small as possible.

It is important to note that the value of \(S\) is heavily influenced by the choice of hyper-parameters, as detailed in Sec.~\ref{sec:details}. In these benchmarks we strived for high accuracy by requiring that both the energy \(\mathcal{E}\) and its gradient \(\mathbf{F}\) are estimated to a precision of at least \(\bar{\Sigma}=10^{-3}\). Furthermore, a minimum of \(2d+1\) training points are required within a grid with tight lattice spacing \(a=0.0125\) for each test point. This ensures that new calculations are invoked when the exploration enters a new parameter region spaced by more than \(a\).

In the forthcoming benchmark calculations, we aim to address three specific scientific questions for evaluating the utility and efficiency of our gGA+AL method:
\begin{enumerate}
    \item \textbf{Ability to learn:} Can a sweep of gGA+AL calculations, once completed and with the data stored, be repeated without requiring any new data for the Hamiltonian parameters already explored? This is a necessary condition for realizing the computational benefits of our data-driven approach. 
    \item \textbf{Transfer-Learning Efficiency:} If multiple sweeps are performed, each with different settings such as \(\Delta U\) or \(t_{ij}\), can data acquired in one sweep be leveraged in another to reduce the need for new calculations? We are interested in whether the explored parameters can span a latent space with overlapping regions that can be exploited for computational efficiency in future calculations.
    \item \textbf{Accuracy Preservation:} Is the accuracy in physical quantities preserved when completing a calculation using the gGA+AL method? While it is always feasible to refine the results through a few standard gGA iterations without active learning at the end of any gGA+AL calculation, achieving high accuracy directly with AL is preferable for maximizing computational gains.

\end{enumerate}

\subsection{Benchmarks for Hubbard Model on Infinite-Coordination Bethe Lattice}\label{sec:benchmarkB}

In this subsection we present benchmarks of our gGA+AL method as applied to the Hubbard model represented by the following Hamiltonian:
\begin{equation}
\label{Hubbard_model}
    \hat{H}=
    \frac{U}{2}\sum_i\left(\hat{n}_i-1\right)^2
    -t\sum_{\langle i,j \rangle}\sum_{\sigma=\uparrow,\downarrow}\left(c^\dagger_{i\sigma}c_{j\sigma}+\text{H.c.}\right),
\end{equation}
where $t$ the hopping between nearest-neighbor sites, and
the the hopping parameters are finite only between nearest-neighbour sites on a infinite-coordination Bethe lattice, with semicircular density of states $\rho(\omega)=2\sqrt{D^2-\epsilon^2}/(\pi D^2)$.

The energy measure is set in unit of the half-bandwidth $D \propto t$.

\subsubsection{Efficiency starting with empty database}

\begin{figure}
    \centering
    \includegraphics[width=0.49\textwidth]{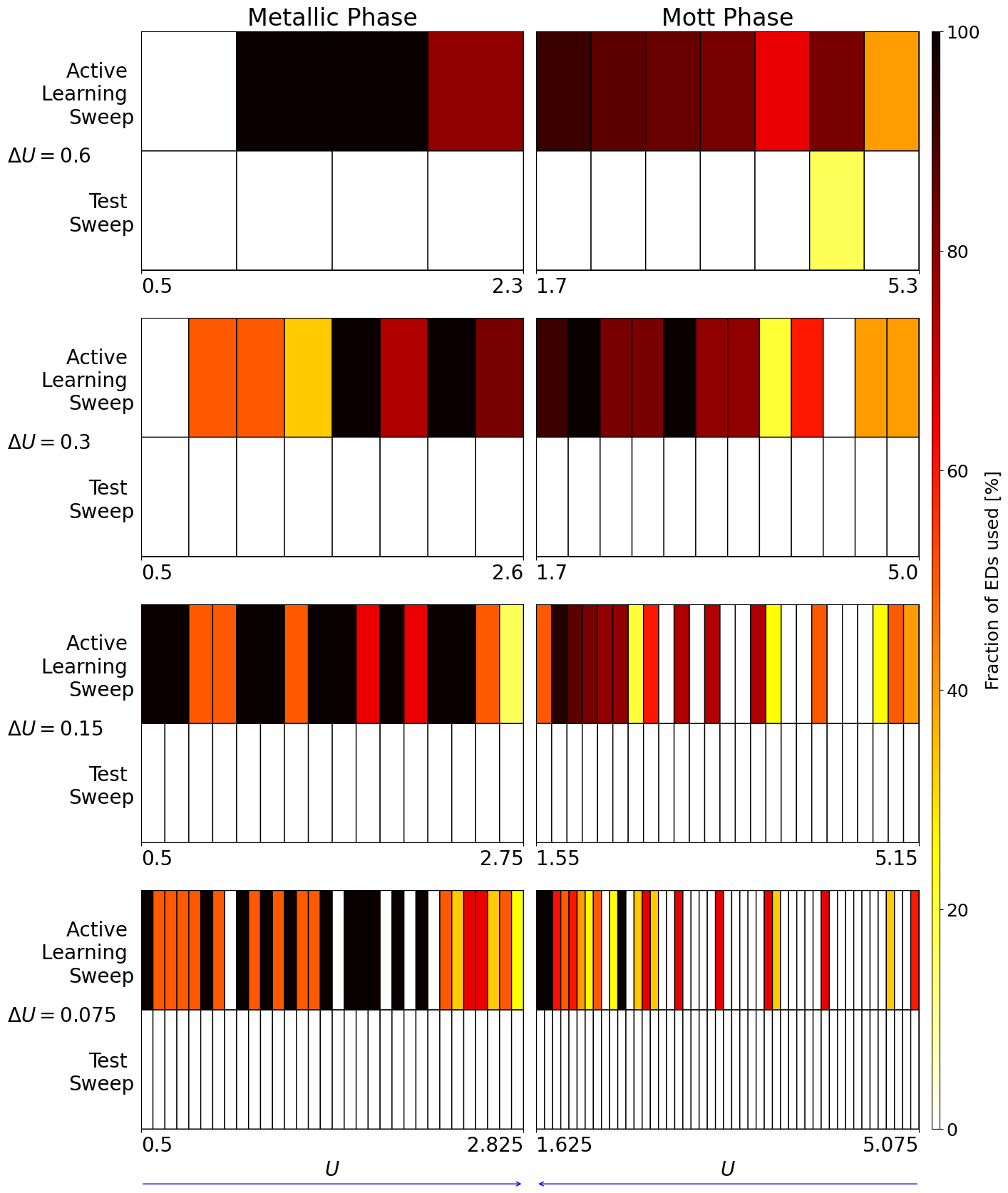}
    \caption{Efficiency metric \(S\) for sweeps performed at different values of \(\Delta U\). Each row corresponds to a unique \(\Delta U\). The left and right columns show \(S\) values for metallic and Mott states, respectively. The results of both the original sweep and the subsequent test sweep are included. 
    }
    \label{Figure2}
\end{figure}

For each \(\Delta U\), we start with an empty database and execute a sweep of calculations, as defined in Section~\ref{sec:benchmarks-goal}. Immediately following this, a test sweep is performed, leveraging the data acquired during the initial sweep. Once the test sweep for a specific \(\Delta U\) is completed, the database is reset to empty, and the entire process is repeated for the subsequent \(\Delta U\) values.
The results of these benchmarks are summarized in Fig.~\ref{Figure2}, where each row corresponds to a different \(\Delta U\). The figure showcases the value of the efficiency metric \(S\) for both the original and test sweeps. The left and right columns of the figure display the $S$ values for metallic and Mott states, respectively.

A key observation is that no additional training data are required in the test sweeps for all \(\Delta U\) values, with the only exception at \(\Delta U=0.6\), where a few new training data are added to the database. 
This confirms the ability of our framework to "learn monotonically," in the sense outlined in Section~\ref{sec:benchmarks-goal}. 
It is also remarkable that the computational gains achieved through our gGA+AL approach are substantial even in the initial sweep, when starting from an empty database. In fact, the average of the efficiency metric $S$ registered throughout each sweep at $\Delta U=0.6,0.3,0.15,0.075$ is at 83\%, 73\%, 56\%, and 40\%, respectively, when considering both the Mott and metallic phases.
The fact that tighter meshes lead to increased overall computational savings is explained by the fact that data acquired along the way for solving the gGA equations can be used by the AL framework for reducing computational cost of subsequent calculations with similar interaction strengths.

From the physical perspective, a very interesting feature of the results shown in Fig.~\ref{Figure2} is that \(S\) is smaller in the Mott phase and the strongly-correlated metallic phase, pointing to higher transfer-learning efficiency in these regions, compared to the weakly correlated regime.
This finding, and its physical interpretation, is discussed later in Sec.~\ref{sec:PCA} with a PCA.

\subsubsection{Efficiency with Progressive Data Accumulation}\label{sec:progressive-Bethe}

\begin{figure}
    \centering
    \includegraphics[width=0.49\textwidth]{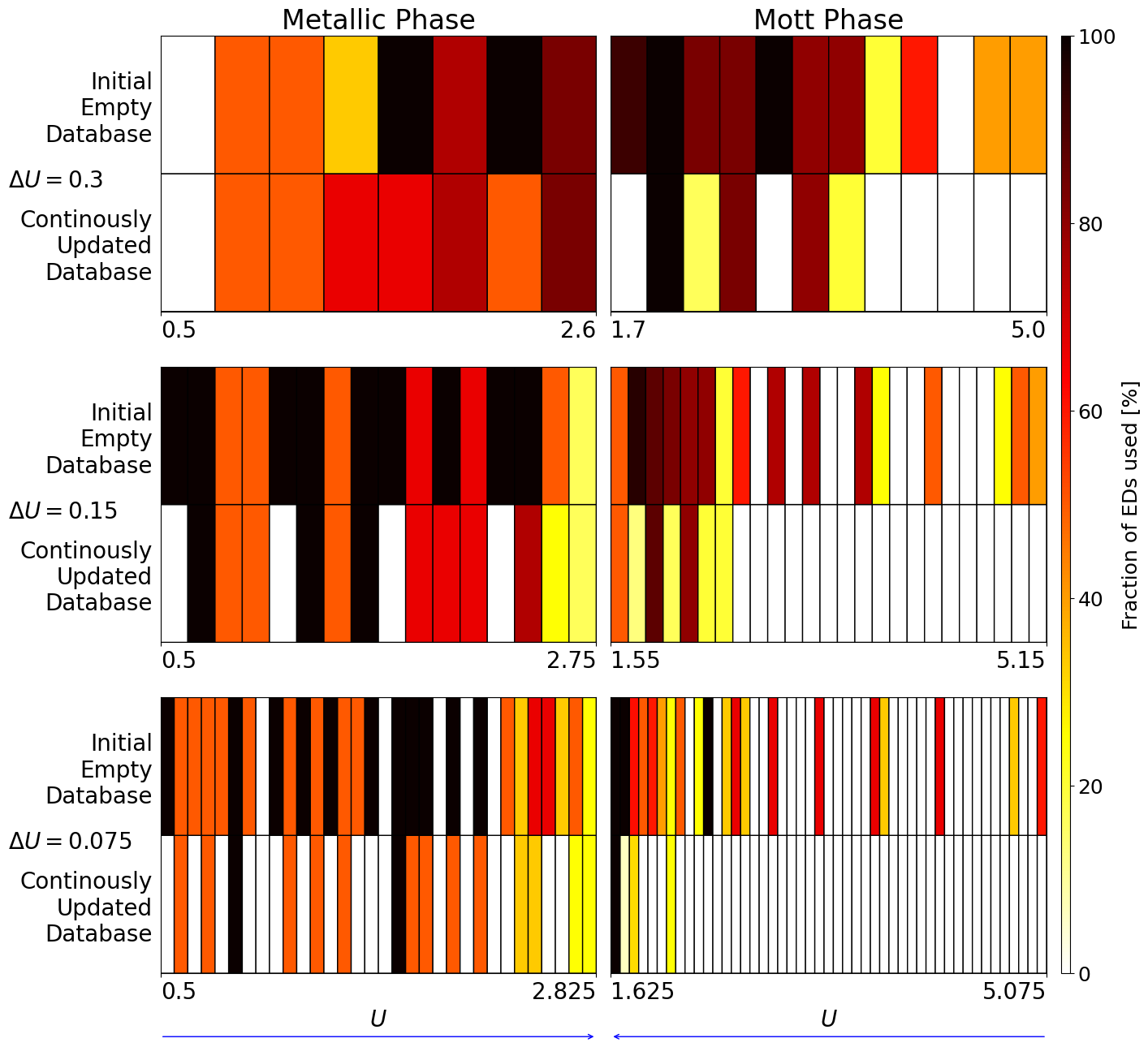}
    \caption{Comparison of the efficiency metric \(S\) for different sweeps: \(\Delta U = 0.3, 0.15, 0.075\). Each row corresponds to mesh spacing \(\Delta U\). The left and right columns represent the \(S\) values for metallic and Mott states, respectively. For each \(\Delta U\), the top panel shows \(S\) calculated using an empty database at the start of each sweep. The bottom panel displays \(S\) calculated starting from the database obtained at the end of the \(\Delta U = 0.6\) sweep and updated continuously as we proceed through the series of \(\Delta U\), from the largest to the tightest.}
    \label{Figure3}
\end{figure}

In Fig.~\ref{Figure3}, we compare the efficiency of our gGA+AL method across the sweeps with spacings \(\Delta U = 0.3, 0.15, 0.075\). Each row of the figure corresponds to one of such sweeps. The top panels present the \(S\) values obtained when starting with an empty database for each new sweep, as in Fig.~\ref{Figure2}. In contrast, the bottom panels show the \(S\) values calculated starting from a database that was initially populated at the end of a \(\Delta U = 0.6\) sweep and subsequently updated without resetting as we traverse through the mentioned series of \(\Delta U\) values.

We observe that, using the scheme that retains data, the metric \(S\) demonstrates a computational gain of over 50\% compared to calculations performed with a reset database. Specifically, in this same scheme, the average of the efficiency metric \( S \) registers at 33\%, 22\%, and 10\% for \( \Delta U = 0.3, 0.15, 0.075 \), respectively.
Consistently with the trend of the results in Fig.~\ref{Figure2}, this gain is even more significant in the regime of strongly-correlated parameters. 
This further supports the transfer-learning ability of our AL framework.

It is important to note that, as mentioned in Sec.~\ref{sec:red-comp}, in the gGA framework the sequence of points $\bX_i$ explored throughout the computation of any given model always converges towards the point $\bar{\bX}$ realizing the corresponding solution. 
Thus, the majority of these points, where \(\E(\bX)\) needs to be learned, tend to cluster around \(\bar{\bX}\) instead of being distributed throughout the entire parameter space.
As a result, the data effectively form an approximate 1-dimensional latent structure (parametrized by $U$), embedded within the 6-dimensional ambient space, as we are going to show explicitly in Sec.~\ref{sec:PCA} with a PCA.
The "transfer-learning" ability displayed by our AL framework (i.e., its ability to exploit the data previously stored for improving efficiency of subsequent calculations) is grounded on its ability to exploit such type of low-dimensional structures, effectively bypassing the exponential computational burden associated with the unnecessary task of learning \( \E(\bX) \) in the whole ambient space.

\subsection{Benchmarks for Hubbard Model with different lattice structures}\label{sec:benchmarkC}

Here we extend our analysis to consider different geometries, specifically variations in the hopping matrices \( t_{ij} \). This enables us to evaluate how the gGA+AL framework performs when the data do not naturally span an approximately 1-dimensional curve solely parametrized by \( U \). A central question we aim to address is whether a low-dimensional structure is commonly present among the embedding parameters \( \bX \) in these more general scenarios, and if so, whether our AL framework can leverage this structure for more efficient and accurate calculations.

\subsubsection{Efficiency with Progressive Data Accumulation}

\begin{figure}
    \centering
    \includegraphics[width=0.49\textwidth]{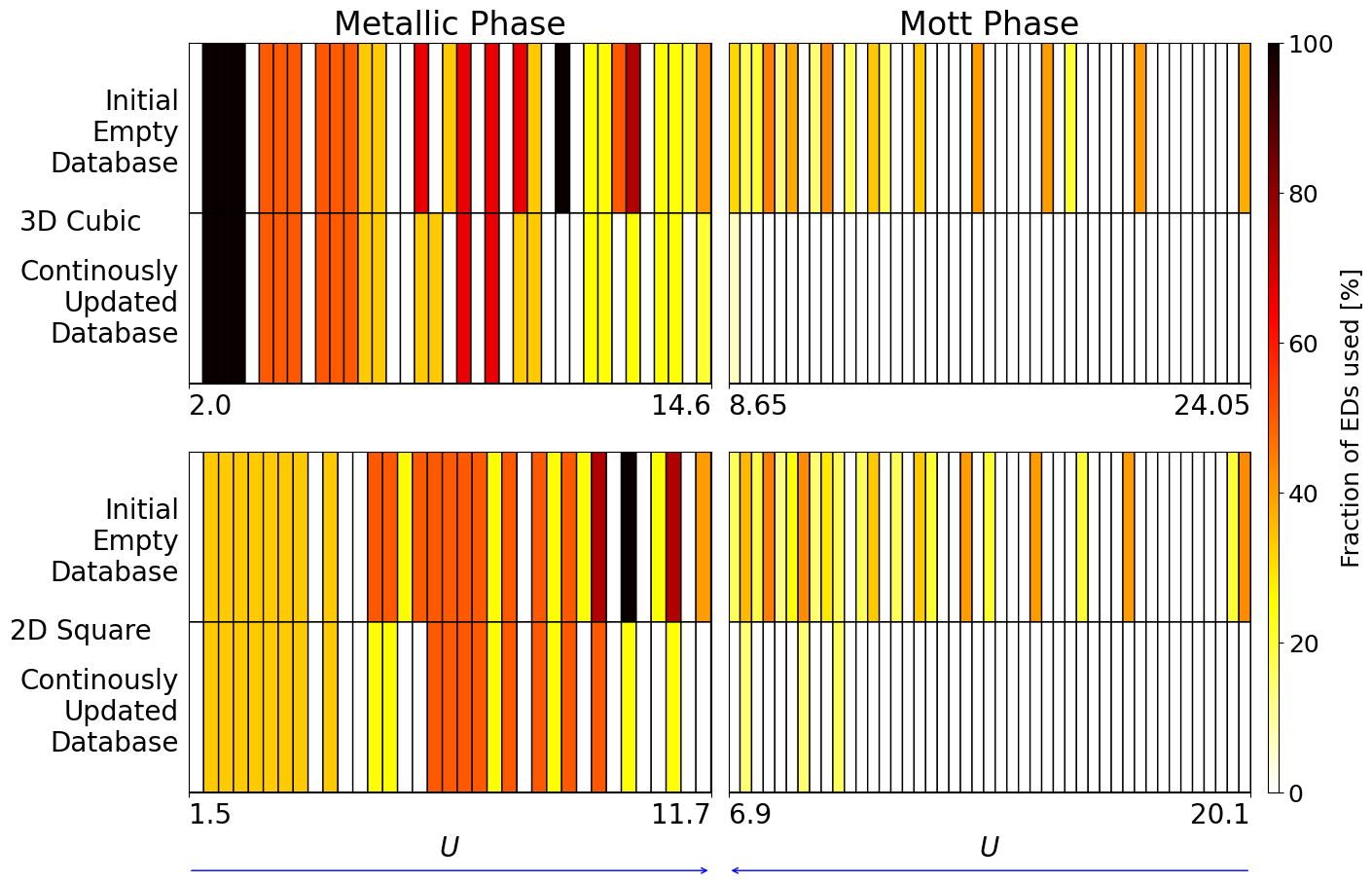}
    \caption{Efficiency metric \( S \) for 3D cubic (upper panels) and 2D square (lower panels) lattices. Each panel is divided into two sections: The upper section shows results obtained without storing any data, while the lower section presents results where all previously acquired data, including that from the Bethe lattice calculations, were retained. Mott points are depicted on the right side, and metallic points are on the left.}
    \label{Figure4}
\end{figure}

To extend the scope of our analysis, we have also performed calculations of the Hubbard model on 3D cubic and 2D square lattices, employing a mesh with \( \Delta U = 0.075 \). The resulting efficiency metrics \( S \) for these calculations are illustrated in Fig.~\ref{Figure4}. 
\begin{figure*}
    \centering
    \includegraphics[width=0.98\textwidth]{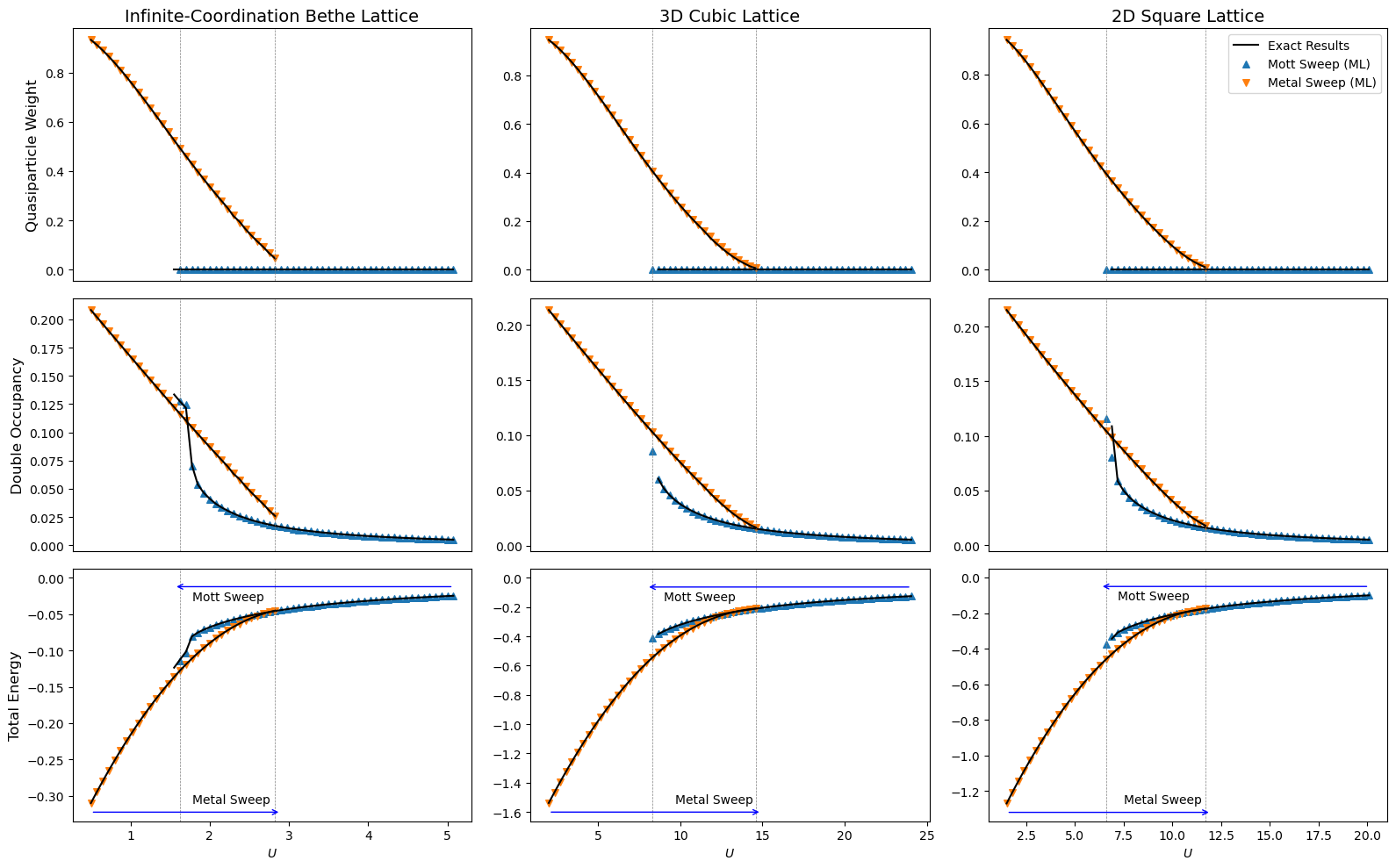}
    \caption{Comparison of total energy, local double occupancy, and quasi-particle weight calculated using gGA+AL and standard gGA methods. The results referred to as "exact results" correspond to the exact gGA solutions and are represented by continuous black lines. The results obtained with gGA+AL are represented by triangles.}
    \label{Figure5}
\end{figure*}

The figure is organized as follows: the upper panels correspond to the 3D cubic lattice calculations, while the lower panels are for the 2D square lattice. Within each panel, the upper and lower parts distinguish between the two modes of data acquisition. The upper part of each panel displays the efficiency metrics obtained without any stored data, whereas the lower part showcases results when retaining all previously acquired data, including that from our Bethe lattice calculations. 

The results of Fig.~\ref{Figure4} are consistent with our previous findings. Even when starting from an empty database, there is a significant computational advantage. More notably, employing the data acquisition model that continuously accumulates data results in additional gain compared to the data-reset calculations. This suggests that despite the differences in geometry between the systems, there is a degree of overlap in the data that is explored. Intriguingly, these gains are predominantly observed in the Mott phase and strongly correlated regime, implying a greater degree of overlap in these cases compared to the weakly correlated regime.

These observations lead us to conclude that the latent space of the "physically relevant embeddings," or the set of parameters \( \bX \) probed during gGA ground-state calculations, possesses a "special" structure. Specifically, this structure may be such that it occupies only a small subset of the ambient parameter space.
We delve deeper into understanding this "special" structure of the latent space in Sec.~\ref{sec:PCA}, where we employ the PCA to study the data structure in detail and provide a physical interpretation of these findings.

\subsubsection{Accuracy of gGA+AL solution}

In addition to computational efficiency, another critical aspect of our gGA+AL framework is its accuracy in calculating physical observables. To rigorously evaluate this, we consider observables such as the total energy, local double occupancy, and quasi-particle weight. These observables are computed from the variational parameters obtained after convergence, as detailed in Sec.~\ref{sec:gauge}.

From Fig.~\ref{Figure5}, it is evident that the application of our ML algorithm does not result in a significant loss of accuracy compared to a canonical gGA algorithm. Also, the endpoint of the metal-insulator transition \( U_{c1} \) is in perfect agreement with that obtained using the standard method. The only discrepancy is that the endpoint of the metal-insulator coexistence region \( U_{c2} \) shows a slight overestimation on the 2D square lattice when using our ML algorithm.

It is also worth pointing out that, as previously mentioned in Sec.~\ref{sec:benchmarks-goal}, it is always possible to refine the results with a few iterations of the standard gGA method after the active learning steps. This ensures that the computational efficiency gained by the gGA+AL framework does not compromise accuracy, providing a risk-free framework for high-efficiency, reliable calculations.

\begin{figure}[ht]
    \centering
    \includegraphics[width=0.52\textwidth]{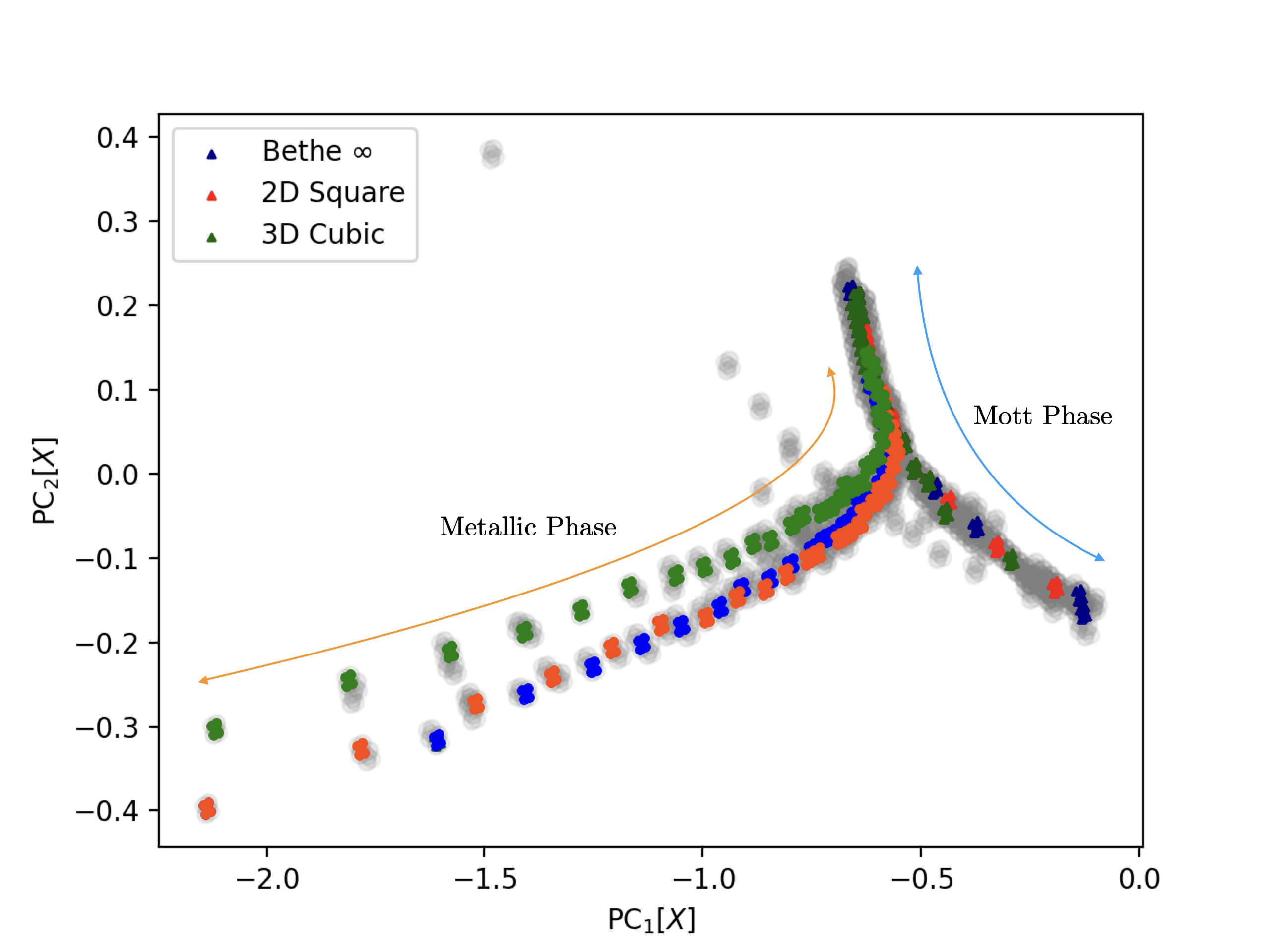}
    \caption{Scatter plot of the first two principal components of the training database. Points obtained after convergence for the Bethe lattice, 2D square lattice, and 3D cubic lattice are colored in blue, red, and green, respectively. All other points are colored in grey.}
    \label{Figure6}
\end{figure}

\subsection{PCA analysis of the training database}\label{sec:PCA}

In this section, we turn our attention to the underlying structure of the database that has been acquired in the course of our calculations. Our primary objective is to probe the latent space within which our AL framework operates. In particular, we aim to elucidate why our AL framework shows notably higher transfer-learning efficiency in the strongly-correlated regime, and to discuss the physical implications of these findings.

Our database comprises vectors \( \bX_\alpha \) from calculations on the Bethe lattice in the limit of infinite coordination number, the 2D square lattice, and the 3D cubic lattice, all initiated without pre-existing data. 
To investigate the low-dimensional structure of such "latent space"
of embedding parameters, we perform a PCA analysis.

\subsubsection{Definition of the PCA}

The PCA analysis of our database consists of the following steps.
\begin{itemize}
\item We construct a $N\times d$ data matrix \( \mathbf{M} \), where
$d=6$ is the dimension of the vectors $\bX_\alpha$ and $N$ is the number of database points,
by placing each EH parameter vector \( \bX_\alpha \) of the database as a row, leading to:
\begin{equation}
\mathbf{M} = 
\begin{bmatrix}
\vdots \\
\bX_\alpha \\
\vdots
\end{bmatrix}
\,,
\end{equation}
i.e., $\mathbf{M}_{\alpha i}=[\bX_\alpha]_i$.

\item The Singular Value Decomposition (SVD) of \( \mathbf{M} \) is represented as a sum of the outer products of its singular vectors:
\begin{equation}
\mathbf{M} = \sum_{r=1}^{d} \sigma_r \mathbf{u}_r \mathbf{v}_r^\mathrm{T}
\,,
\label{PCA-exact}
\end{equation}
where \( \sigma_r \) denotes the singular values sorted from the largest to the smallest,
the vectors \( \mathbf{u}_r \) and \( \mathbf{v}_r \) are the column vectors corresponding to the left and right singular vectors, respectively. 
From Eq.~\eqref{PCA-exact} it follows that the data points $\bX_\alpha$ can be expressed as the following expansion of $\mathbf{v}_r^\mathrm{T}$:
\begin{equation}
    \bX_\alpha=\sum_{r=1}^{d} \sigma_r 
    [\mathbf{u}_r]_\alpha \mathbf{v}_r^\mathrm{T}
    \,,
    \label{PCA-X-exact}
\end{equation}
with coefficients $x_{r\alpha}=\sigma_r 
    [\mathbf{u}_r]_\alpha$,
where $[\mathbf{u}_r]_\alpha$ denotes the $\alpha$ component of  $\mathbf{u}_r$.
Since the singular values $\sigma_r$ are sorted from the largest to the smallest, the first terms of Eq.~\eqref{PCA-X-exact} retain the most significant features of the parameter space.

\item 
The approximation of each data point \( \bX_\alpha \) is thus obtained by truncating the sum in Eq.~\eqref{PCA-X-exact} as follows:
\begin{equation}
    \bX_\alpha=\sum_{r=1}^{d_{\text{cut}}} \sigma_r 
    [\mathbf{u}_r]_\alpha \mathbf{v}_r^\mathrm{T}
    \,,
    \label{PCA-X-exact2}
\end{equation}
where \( d_{\text{cut}} \) is the number of retained principal components selected to capture the desired amount of total variance from \( \mathbf{M} \),
and $\mathbf{v}_r^\mathrm{T}$ are the corresponding "principal axes". 
\end{itemize}


\subsubsection{Application of the PCA}

The results of the PCA analysis described above are in Fig.~\ref{Figure6},
which shows the first two principal components, i.e. $[\mathbf{u}_1]_\alpha\sigma_1$ and $[\mathbf{u}_2]_\alpha\sigma_2$, where $\sigma_1=77.8$ and $\sigma_1=14.6$. These two principal components account for more than 88 \% of the variability of the data.

To further investigate the low-dimensional structure of the latent space, we present a scatter plot of these first two principal components in Fig.~\ref{Figure6}, providing us with a pictorial representation of the latent space. In this plot, the points are color-coded based on the lattice type and the stage of the calculation. Specifically, points obtained after convergence for the Bethe lattice, 2D square lattice, and 3D cubic lattice are colored in blue, red, and green, respectively. All other points, which are gathered during the self-consistency procedure but do not correspond to converged solutions, are colored in grey.

In line with our earlier discussion in Sec.~\ref{sec:progressive-Bethe}, the data for each lattice effectively form an approximate one-dimensional latent curve, parametrized by \( U \), which bifurcates within the coexistence region. 
Remarkably, data subsets corresponding to each lattice structure are very similar. Furthermore, we observe that the separation between the data corresponding to different lattices is more pronounced in the weakly correlated regime (small \( U \), lower-left part of the graph). As \( U \) increases, the data corresponding to different lattices in the metallic regime become increasingly overlapping, culminating in maximum overlap near the end of the coexistence region \( U_{c2} \). 

From the computational perspective, these observations shed light on the higher transfer-learning efficiency of our AL framework in the strongly correlated regime. The overlapping data imply that similar regions of the feature space are explored across different calculations. Consequently, the data from one calculation can be effectively transferred to subsequent calculations, reducing the need for additional data points.

From a physical standpoint, the observed overlapping behavior of the databases across different lattice structures is rooted in the universality of Mott physics. The parameters \( \bX_\alpha \) obtained after convergence, represented by the colored dots in Fig.~\ref{Figure6}, can be interpreted as physical embeddings of the correlated fragments. These embeddings capture the essence of electron localization induced by the Hubbard interaction, which transcends the specifics of the lattice structure. As we approach and enter the Mott phase in the strongly correlated regime, the fragments become less entangled with their surrounding environment due to reduced charge fluctuations. Consequently, it is understandable that the physical embeddings become increasingly less dependent on the lattice type, leading to the observed overlaps in the databases.

\subsubsection*{Possible Future Methodological Enhancements}

In light of our findings concerning the low-dimensional latent space, and their general origin rooted in Mott physics, it is natural to consider additional computational techniques that could further leverage this structure in future applications to complex multi-orbital strongly correlated systems. 
Specifically, Deep Kernel Learning (DKL) with autoencoders could further facilitate learning within our AL framework, as it could offer enhanced flexibility and scalability for discovering optimal feature spaces, all while preserving the essential element of uncertainty quantification employed within our AL procedure.

\section{Conclusions and Outlook}

In this study, we have presented an AL framework integrated within the gGA to efficiently explore the ground state of the EH in the context of the single-band Hubbard model. From a computational standpoint, this approach leads to a marked reduction in the number of EH instances that must be solved iteratively, thus significantly mitigating the computational cost inherent to gGA. Moreover, our PCA analysis reveals that the parameters of the EH reside in a latent space with a low-dimensional structure that is largely invariant to the specifics of the lattice geometry, especially in the strongly-correlated regime.

From the physical perspective, the existence of this low-dimensional latent space can be attributed to the universal features of Mott physics. The phenomenon of electron localization, caused by the Hubbard interaction, transcends the geometric specifics of the systems studied. As a result, the correlated fragments are less susceptible to environmental influences, leading to a latent space whose characteristics are conserved across various lattice structures.

Looking forward, extending this methodology to more complex systems involving multiple orbitals, such as 5-orbital d-systems and 7-orbital f-systems, presents an interesting challenge. While the universality of Mott physics gives us reason to expect similar low-dimensional structures in these more complicated systems, the actual existence and dimensionality of such a latent space remains an open question. 
In this respect, it is important to note that, when considering real-material calculations, the parameters of the electronic Hamiltonian are not freely adjustable. Structural stability, which emerges from the interplay between electronic and lattice degrees of freedom, imposes further constraints on the physically realizable electronic structures, which do not exist in model calculations, where all parameters can be tuned in arbitrary ways. 
A trivial example of how structural stability limits the possible quantum embeddings of the correlated degrees of freedom is that it often leads to symmetry, which can be exploited to reduce the number of gGA parameters using group-theoretical considerations. 
Additionally, structural stability restricts the possible atomic environments based on fundamental principles of chemistry, such as valence compatibility between atoms. These constraints may significantly limit the dimensionality and structure of the latent space of physically realizable embeddings, facilitating the learning problem. 

Hence, the implementation of our AL framework in real-material calculations, potentially within an ab-initio DFT+gGA framework, could provide further insights into the structure of this latent space and its limitations, laying the groundwork for more efficient simulations of complex strongly-correlated materials.

\begin{acknowledgments}

This work was supported by a grant from the Simons Foundation (1030691, NL). 
We gratefully acknowledge funding from the Novo Nordisk Foundation
through the Exploratory Interdisciplinary Synergy Programme project
NNF19OC0057790.
DGA acknowledges funding from the Rising Star Fellowship of the Department of Biology, Chemistry, Pharmacy of Freie Universit\"at Berlin.
The part of the work by YXY was supported by the U.S. Department of Energy (DOE), Office of Science, Basic Energy Sciences, Materials Science and Engineering Division, including the grant of computer time at the National Energy Research Scientific Computing Center (NERSC) in Berkeley, California. This part of research was performed at the Ames National Laboratory, which is operated for the U.S. DOE by Iowa State University under Contract No. DE-AC02-07CH11358. KB acknowledges support from the LANL LDRD Program.

\end{acknowledgments}


%

\newaliascnt{suppeqn}{equation}
\let\oldtheequation\theequation
\let\theequation\thesuppeqn

\clearpage


\onecolumngrid
\begin{center}
\textbf{\large Supplemental Information for: \\
Active Learning approach to simulations of Strongly Correlated Matter with the
Ghost Gutzwiller Approximation}
\vspace{0.4cm}
\end{center}

\setcounter{equation}{0}
\setcounter{figure}{0}
\setcounter{section}{0}
\setcounter{table}{0}
\setcounter{page}{1}
\makeatletter

\newpage

\maketitle

\section{Gaussian Process Regression in the Context of Active Learning}

Within our active learning (AL) algorithm described in the main text, we store data gathered during the ghost Gutzwiller Approximation (gGA) iterations in a training database
\begin{align} 
\label{db-SM}
    \mathbf{D} = \{\mathcal{E}(\bX), \mathbf{F}(\bX), \bX, \sigma_0, \pmb{\sigma}\}
\end{align}
consisting of all the quantities needed for training our ML model and making predictions. In Eq.~\eqref{db-SM} $\E(\bX)$ and $\mathbf{F}(\bX)$ refer to the energy and its $d$ gradient components evaluated at a point $\bX$ in a $d$ dimensional space. Furthermore, the model's prior uncertainties $\sigma_0$ and $\pmb{\sigma}$ for the energy and its gradients are stored. These uncertainties take into account noise in the training and test data. However, since in our case all the training and test data is deterministic, they only are used as regularization parameters and, hence, we use $\sigma_0 = \sigma_i$ $\forall\,i \in \{1, 2, \dots d\}$.

During the gGA iterations we encounter new points $\bX^*$ for which we do not know the corresponding energies and gradients. For such a new test point we then attempt to predict its energy and gradient using Gaussian process regression (GPR). If there are closeby points (nearest neighbors) available in the database $\mathbf{D}$, their energies and gradients are retrieved from the database and used as training points for GPR. All of the remaining nearest neighbors to $\bX^*$ which are not yet available in the database are then generated on the fly and their energies and gradients are merged into the database until we have the full set of required $2d + 1$ training  points required by our algorithm (for more details on the hypercubic lattice used for discretizing the space of training points it is referred to the main text).

This local subset of $\mathbf{D}$ is stored in a vector
\begin{align}
    \mathbf{y}(\bX_i) = \left(\mathcal{E}(\bX_i), F_1(\bX), F_2(\bX_i), \dots, F_{d}(\bX_i)\right)^\mathrm{T}
\end{align}
of length $N(d+1)$ containing the energy and its $d$ gradients for all of the $N=2d+1$ training points, where the superscript "$\mathrm{T}$" denotes the transpose. This training data will be used to define equations for making prediction at the test point $\bX^*$.

\subsection{Gradient Enhanced GPR}

Using our prior knowledge about the data given by the training data under the assumption that all elements of $\mathbf{y}(\bX_i)$ for any given point $\bX_i$ are normally distributed we arrive at the following equations for the posterior mean and covariance matrix
\begin{align} \label{eq:g_posterior_mean}
    \bar{\mathbf{y}} &= (\mathbf{K}^*)^{\mathrm{T}} (\mathbf{K} + \sigma^2 {\bf{I}})^{-1}  \mathbf{y}(\bX)
    \\
    {\pmb{\Sigma}} &= 
     \mathbf{K}^{**} -  (\mathbf{K}^*)^{\mathrm{T}} (\mathbf{K} + \sigma^2 {\bf{I}})^{-1} \mathbf{K}^{*},
\end{align}
where $\bar{\mathbf{y}}$ is the posterior mean vector of length $N^*(d+1)$ containing the expectation values (predictions) for the energy and its gradients for the $N^*$ test points $\bX_i^*$, i.e.
\begin{align}
    \bar{\mathbf{y}}(\bX_i^*) = \left(\langle\mathcal{E}(\bX_i^*)\rangle, \langle F_1(\bX_i^*)\rangle, \langle F_2(\bX_i^*)\rangle, \dots, \langle F_{D}(\bX_i^*)\rangle
    \right)
\end{align}
and $\mathbf{K}$, $\mathbf{K}^*$ and $\mathbf{K}^{**}$ are the modified prior covariance matrices of size $N(d+1)\times N(d+1)$, $N(d+1)\times N^*(d+1)$ and $N^*(d+1)\times N^*(d+1)$, respectively. The modified structure of these covariance matrices ensures that the predicted energy and its gradients are consistent, i.e. $\langle\mathbf{F}\rangle = \nabla\langle\mathcal{E}(\bX)\rangle$, by also incorporating the gradients of the kernel function as follows:

\begin{equation} \label{g_covar}
  {\mathbf{K}} =
  \left( {\begin{array}{ccccccc}
   k^{0,0}_{1,1} & {\mathbf{k}}^{0,1}_{1,1} &
   k^{0,0}_{1,2} & {\mathbf{k}}^{0,1}_{1,2} &
   \cdots
   & k^{0,0}_{1,N} & {\mathbf{k}}^{0,1}_{1,N}  \\
   {\mathbf{k}}^{1,0}_{1,1} & {\mathbf{k}}^{1,1}_{1,1}  &
   {\mathbf{k}}^{1,0}_{1,2} & {\mathbf{k}}^{1,1}_{1,2}    &
   \cdots &
   {\mathbf{k}}^{1,0}_{1,N} & {\mathbf{k}}^{1,1}_{1,N}  \\

      \vdots      &    \vdots     &  \vdots      &    \vdots    &   \ddots     &   \vdots    &   \vdots    \\
   k^{0,0}_{N,1} & {\mathbf{k}}^{0,1}_{N,1} &  
   k^{0,0}_{N,2} & {\mathbf{k}}^{0,1}_{N,2} & 
   \cdots    & 
   k^{0,0}_{N,N} & {\mathbf{k}}^{0,1}_{N,N} \\
   {\mathbf{k}}^{1,0}_{N,1} & {\mathbf{k}}^{1,1}_{N,1} & 
   {\mathbf{k}}^{1,0}_{N,2} & {\mathbf{k}}^{1,1}_{N,2}   &
   \cdots & 
   {\mathbf{k}}^{1,0}_{N,N} & {\mathbf{k}}^{1,1}_{N,N} \\
  \end{array} } \right),
\end{equation}

Here, the following short-hand notations is introduced for the original values and partial derivatives of the kernel function,
\begin{align}
  k_{i,j}^{0,0} & =
  k\left( {\mathbf{X}}_i, {\mathbf{X}}_j \right)
  \,,
\\
  \mathbf{k}_{i,j}^{1,0} &= \nabla^\mathrm{T}_{\mathbf{X}_i} k\left( {\mathbf{X}}_i, {\mathbf{X}}_j \right)
  \,,
\\
  \mathbf{k}_{i,j}^{0,1} & 
  =\nabla_{\mathbf{X}_j} k\left( {\mathbf{X}}_i, {\mathbf{X}}_j \right)  
  \,,
\\
  \mathbf{k}_{i,j}^{1,1} &
  =\nabla^\mathrm{T}_{\mathbf{X}_i} \nabla_{\mathbf{X}_j} k\left( {\mathbf{X}}_i, {\mathbf{X}}_j \right)
  \,.
\end{align}
where the kernel function used for this work (square exponential kernel) is given by 
\begin{align} \label{kernel}
    k(\mathbf{X}_i, \mathbf{X}_j) = \sigma_f^2 \exp \left( - \frac{(\mathbf{X}_i - \mathbf{X}_j )^2}{2l^2}   \right)\,.
\end{align}

This also allows for the computation of standard deviations (uncertainties) of energies and gradients for a test point $\bX_i^*$
\begin{align} \label{g_variance0}
    \Sigma^i_{0} &= \left[\langle\mathcal{E}^2(\bX_i^*)\rangle - \langle\mathcal{E}(\bX_i^*)\rangle^2\right]^{\frac{1}{2}}
    \\ \label{g_variance1}
    \Sigma^i_{j} &= \left[\langle F_j^2(\bX_i^*)\rangle - \langle F_j(\bX_i^*)\rangle^2 \right]^{\frac{1}{2}}
    \,.
\end{align}

By utilizing this modified structure of the covariance matrix, the GPR model effectively leverages both function and gradient observations, often yielding enhanced predictions while at the same time allowing us to learn a scalar function and being able to use the gradient predictions for the construction of the density function. For more information on the program implementation of this approach as well as on the detailed analysis of its computational scaling, the interested reader is referred to Refs.~\cite{schmitz2020-SM,arti2023-SM}.

\section{The $\text{g}$GA Approximation}

The goal of this section is to outline the algorithmic structure of the gGA.

\subsection{The gGA Lagrange function}

As explained in the main text, for the single-band models of the form considered in this work,
the gGA framework is encoded the following Lagrange function:
\begin{align}
&\Lag[
{\Phi},E^c;\,  \R,\Lambda;\, \D, \Lambda^{c};\,\Delta, \Psi_0, E;\mu]=
\nonumber\\&\;
= \Av{\Psi_0}{\h_{\text{qp}}[\R,\Lambda]}
+E\left(1-\langle\Psi_0|\Psi_0\rangle\right)
\nonumber\\&\;
+\sum_{i=1}^{\mathcal{N}}
\left[\Av{\Phi_i}{\h^i_{\text{emb}}[\D_i,\Lambda_i^c,U,\mu]}
+E_i^c\left(1-\langle \Phi_i | \Phi_i \rangle
\right)\right]
\nonumber\\&\,
-\sum_{i=1}^{\mathcal{N}}\left[
\sum_{\sigma=\uparrow,\downarrow}\sum_{a,b=1}^{B}\big(
[\Lambda_i]_{ab}+[\Lambda^c_i]_{ab}\big)[\Delta_i]_{ab}
\right.
\nonumber\\&\left.
\qquad+\sum_{\sigma=\uparrow,\downarrow}
\sum_{c,a=1}^{B}
\big(
[\D_{i}]_{a} [\R_{i}]_{c}
\left[\Delta_i(\mathbb{I}-\Delta_i)\right]^{\frac{1}{2}}_{ca}
+\text{c.c.}\big)
\right]\,,
\label{Lag-SB-emb-SM}
\end{align}
where \(\mathcal{N}\) denotes the total number of unit cells, \(E\) and \(E_i^c\) are scalars, and \(\Delta_i\), \(\Lambda_i\), and \(\Lambda_i^c\) are \(B\times B\) Hermitian matrices. Additionally, \(\D_i\) and \(\R_i\) are \(B \times 1\) column matrices.
The so-called "quasiparticle Hamiltonian" (\(\h_{\text{qp}}\)) and EH (\(\h^i_{\text{emb}}\)) are defined as:
\begin{align}
    \hat{H}_{\text{qp}}[\R,\Lambda]&=\sum_{i=1}^{\mathcal{N}}\sum_{a,b=1}^{{B}}
    \sum_{\sigma=\uparrow,\downarrow}
    [\Lambda_i]_{ab}\,\fc_{ia\sigma}\fa_{ib\sigma}
    \nonumber\\
    &+\sum_{{i,j=1 \atop i \neq j}}^{N} \sum_{a,b=1}^{B}\sum_{\sigma=\uparrow,\downarrow}[\R_i t_{ij}\R_j^\dagger]_{ab}\,\fc_{ia\sigma}\fa_{jb\sigma}
    \,,
    \label{Hqp-SM}
    \\
    \h^i_{\text{emb}}[\D_i,\Lambda_i^c,U, \mu]&= 
    \frac{U}{2}\left(\hat{n}_i-1\right)^2
    -\mu\, \hat{n}_i
    \nonumber\\
    &+ \sum_{a=1}^{B}\sum_{\sigma=\uparrow,\downarrow}\left[[\D_i]_{a}\,{c}^{\dagger}_{i\sigma}{b}^{\phantom{\dagger}}_{ia\sigma}
    +\text{H.c.}\right]\nonumber\\  &+\sum_{a,b=1}^{B}\sum_{\sigma=\uparrow,\downarrow}[\Lambda^c_i]_{ab}\, {b}^{\phantom{\dagger}}_{ib\sigma}{b}^{\dagger}_{ia\sigma}
   \,,
   \label{hemb-SM}
\end{align}
where $\hat{n}_i = \sum_{\sigma} \cc_{i\sigma}\ca_{i\sigma}$.
The integer number \(B\) controls the size of the gGA variational space and, in turn, the precision of the gGA approach.

Below we write explicitly the saddle-point equations obtained by extremizing the Lagrange function above.

\subsection{The gGA Lagrange Equations}
In order to simplify the final saddle point equations for the gGA Lagrangian, we rewrite Eq.~\eqref{Hqp-SM} as follows:
\begin{align}
    \hat{H}_*[\R,\Lambda]
    &=\sum_{i,j=1}^{\mathcal{N}}\sum_{\sigma=\uparrow,\downarrow}
    [\Pi_i h_* \Pi_j]_{ab}\,\fc_{i\sigma}\fa_{j\sigma}
    \,,
    \label{SM-Hqp2}
\end{align}
where we have introduced the matrix:
\begin{align}
    h_* = \begin{pmatrix}
    \Lambda_1  &\R_1t_{12}\R^\dagger_2& \dots & \R_{1}t_{1 \mathcal{N}}\R^\dagger_{\mathcal{N}} \\
    \R_2t_{21}\R^\dagger_1 & \Lambda_2 & \dots & \vdots \\
    \vdots  & \vdots  & \ddots & \vdots \\
    \R_{\mathcal{N}}t_{\mathcal{N} 1}\R^\dagger_{1} & \dots& \dots & \Lambda_{\mathcal{N}}
  \end{pmatrix}
  \label{SM-h*}
\end{align}
and the projectors over the degrees of freedom corresponding to each fragment:
\begin{align}
   \label{SM-Proj}
   \Pi_i = \begin{pmatrix}
     \delta_{i1} & \dots & \mathbf{0} \\
     \vdots  & \ddots & \vdots \\
    \mathbf{0}  &  \dots & \delta_{iM}
  \end{pmatrix}
  \,.
\end{align}

Additionally, we introduce the following parametrization of the Lagrange multipliers $\Delta_i$, $\Lambda_i$ and $\Lambda^c_i$, where they are expanded in terms of a basis of orthonormal matrices $\left[h_i\right]_s$ with respect to the canonical scalar product $(A, B) = \mathrm{Tr}\left[ A^\dagger B\right]$:
\begin{align}
    \label{coeffDelta}
    \Delta_i =& \sum_{s=1}^{B^2} \left[d^0_i\right]_s \left[h^\mathrm{T}_i\right]_s \\
    \label{coeffL}
    \Lambda_i =& \sum_{s=1}^{B^2} \left[l_i\right]_s \left[h_i\right]_s  \\
    \label{coeffLc}
    \Lambda^c_i =& \sum_{s=1}^{B^2} \left[l^c_i\right]_s \left[h_i\right]_s \,,
\end{align}

Given the definitions above, we write the saddle point equations for the gGA Lagrangian as follows:
\begin{align}
    &\hat{H}_*[\R,\Lambda]\ket{\Psi_0} = E_0\ket{\Psi_0}
    \,,
    \label{SM-Hqp-summary}
    \\
    &[\Delta_i]_{ab} = \Av{\Psi_0}{\fc_{ia\sigma} \fa_{ib\sigma}}
    \,,
    \label{SM-Delta-summary}
    \\    
    & 
    \sum_{c=1}^{B} \left[\D_i\right]_{c}\left[\Delta_i\left(\mathbb{I}- \Delta_i\right)\right]_{ac}^{\tfrac{1}{2}} = \sum_j \left[t_{ij}\R^{\dagger}_j \Pi_jf\left(h_*\right)\Pi_i\right]_{ a}
    \,,
    \label{detD-SM}
    \\
    &[l^c_i]_{s} = -[l_i]_{s}-\sum_{c,b=1}^{B}\frac{\partial}{\partial \left[d^0_i\right]_s} \left(\left[\Delta_i\left(\mathbb{I}-\Delta_i\right)\right]^{\tfrac{1}{2}}_{cb}\left[\D_i\right]_{b}\left[\R_i\right]_{c} + \mathrm{c.c.}\right)
    \,,
    \label{detLc-SM}
    \\
    &\hat{H}^i_{\mathrm{emb}}(\mathcal{D}_i, \Lambda^c_i, U, \mu) \ket{\Phi_i} = E_i^c (\mathcal{D}_i, \Lambda^c_i, U, \mu) \ket{\Phi_i} 
    \,,
    \label{SM-Hemb-summary}
    \\
    &\left[\Delta_i\right]_{ab} = \bra{\Phi_i}\bba_{ib\sigma}\bbc_{ia\sigma}\ket{\Phi_i} 
    \,,
    \label{detF2-SM}
    \\
    &\sum_{a=1}^{B\nu_i} [\R_{i}]_{a}
    [\Delta_i(\mathbb{I}-\Delta_i)]^{\frac{1}{2}}_{ab} =\Av{\Phi_i}{\cc_{i\sigma}\bba_{ia\sigma}}
    \,,
    \label{detF1-SM}
\end{align}
where $E_0$ is the ground-state eigenvalue of the quasi-particle Hamiltonian, $E^c_i$ is the ground-state eigenvalue of the $i$-th EH, and $f$ is the zero-temperature Fermi function.

As explained in the main text, because of the Helmann-Feynmann theorem:
\begin{align}
    \bra{\Phi_i}\bba_{ib\sigma}\bbc_{ia\sigma}\ket{\Phi_i} 
    &=\frac{\partial\bar{E}^c}{\partial[\Lambda^c_i]_{ab}}
    \\
   \Av{\Phi_i}{\cc_{i\sigma}\bba_{ia\sigma}}
   &=2\frac{\partial\bar{E}^c}{\partial[\D_i]_{a}} 
   \,,
\end{align}
where:
\begin{align}
&\bar{E}^c(\D, \Lambda^c, U,\mu) =  
\left\langle{\h^i_{\text{emb}}[\D,\Lambda^c, U,\mu]}\right\rangle_{\D,\Lambda^c, U,\mu}
\,,
\label{eq:Ec-SM}
\end{align}
$\h^i_{\text{emb}}$ is the EH defined
in Eq.~\eqref{hemb-SM}, and the expectation value is taken with respect to the corresponding half-filled ground state.

\subsection{Iterative procedure to solve the gGA equations}\label{sec:itergGA}

The equations above can be solved with multiple approaches. 
In the benchmark calculations of this work we used the following iterative procedure:
\begin{enumerate}
    \item Given an initial guess for the parameters $\R_i,\Lambda_i$, solve Eq.~\eqref{SM-Hqp-summary} and determine $\Delta_i$ using Eq.~\eqref{SM-Delta-summary}.
    \item Use Eq.~\eqref{detD-SM} to determine $\D_i$ from the parameters above.
    \item Use Eq.~\eqref{detLc-SM} to determine $\Lambda^c_i$ from the parameters above.
  \item Iteratively determine chemical potential $\mu$ such that the number of physical particles in the EH is equal to the total number of particles in fragment $i$. This is done by solving a root problem for the particle number, which involves the following steps in each iteration: 
        \begin{itemize}
          \item Incorporate chemical potential into $\tilde{\Lambda}^c_i$ as $\tilde{\Lambda}^c_i = u_i^\dagger\Lambda^c_iu_i - \mathbb{I}\mu$.
          \item Solve Eq. \eqref{SM-Hemb-summary} using exact diagonalization.
          \item Compute hybridization and bath blocks $\rho_i^{\mathrm{hyb}}$ and $\rho_i^{\mathrm{bath}}$ of density matrix. 
        \end{itemize}

    \item Use Eq.~\eqref{detF1-SM} to determine $\Delta_i$ from the parameters above.
    \item Use Eq.~\eqref{detF2-SM} to determine $\R_i$ from the parameters above.
    \item Use Eqs.~\eqref{SM-Hqp-summary} and \eqref{SM-Delta-summary} to determine $\Lambda_i$.
    \item Restart from the first step using the so-obtained parameters $\R_i,\Lambda_i$, and iterate until self-consistency is reached (i.e., until the initial and $\R_i,\Lambda_i$
    and those obtained after the steps above are equal up to a gauge transformation, within a given accuracy threshold).
\end{enumerate}

\section{Gauge Fixing}
In this section we are going to explain the details of the ML procedure outlined in the main text, where ML is used to circumvent the bottleneck in gGA calculations, i.e. the repeated solution of Eq. \eqref{SM-Hemb-summary} in step 4 in the algoorithm outlined in Sec.~\ref{sec:itergGA}.

In the main text we have established that the energy function $\bar{E}^c$ of the EH can be expressed in a reduced domain where $\Lambda^c$ is diagonal as follows:
\begin{align}
    \bar{E}^c(\D, \Lambda^c, U,\mu) 
    &=U\mathcal{E}(\tilde{\D}_1,..,\tilde{\D}_B,\tilde{\Lambda}^c_{11},..,\tilde{\Lambda}^c_{BB})+\mu
    \\
    &= U\mathcal{E}(\bX) + \mu
    \,,
\end{align}
where $\mathcal{E}$ is the energy of the EH in this reduced domain and the matrix elements of $\tilde{D}$ and $\tilde{\Lambda}^c$ have been arranged into a vector:
\begin{equation}
\bX=(\tilde{\D}_1,..,\tilde{\D}_B,\tilde{\Lambda}^c_{11},..,\tilde{\Lambda}^c_{BB})
\end{equation}
and the elements of the vector $\bX$ are then given by the following transformations:
\begin{align}
    \label{lcparam}
    \tilde{\Lambda}^c 
    &= \frac{1}{U} 
    \left(u^{\dagger}\Lambda^c u
    +\mu \mathbb{I}\right)
    \\
    \label{Dparam}
    \tilde{\D} &= \frac{1}{U}  u^\mathrm{T} \D \,,
\end{align}
where:
$u =  u^{\mathrm{perm}}  u^{\mathrm{phase}}  \left(u^{\mathrm{eigen}}\right)^\mathrm{T}  \in O(B)$, where 
\begin{itemize}
    \item $u^{\mathrm{eigen}}$ transforms to the eigenbasis of $\Lambda^{\mathrm{c}}$.
    \item $u^{\mathrm{phase}}$ fixes the phase of $\tilde{\D}$ such that $\tilde{\D}_i>0$.
    \item $u^{\mathrm{perm}}$ is a permutation matrix, which ensures that $\tilde{\D}_i\ge\tilde{\D}_{i+1}$. 
\end{itemize}
Since the diagonalization of the EH is gauge perserving, the only outputs available are 
\begin{align}
    \frac{\partial\mathcal{E}}{\partial\tilde{\Lambda}^c_{aa}} &= \tilde{\rho}^{\mathrm{bath}}_{aa} \\
        \frac{\partial\mathcal{E}}{\partial\tilde{\D}_a} &= 2\tilde{\rho}^{\mathrm{hyb}}_{a}
        \,.
\end{align}
This means that we can only directly obtain $\rho^{\mathrm{hyb}}$ and the diagonal elements of $\rho^{\mathrm{bath}}$. However, $\rho^{\mathrm{bath}}$ is generally not diagonal in the gauge defined above. Furthermore, the function $\E$ does not provide us with direct access to the double occupancy. 
Below we derive analytical expressions which enable to compute these quantities from $\E(\bX)$, $\mathbf{F}(\bX)$ and $\bX$.

\subsection{Calculation of off-Diagonal Density Matrix Elements of the EH}

To arrive at an expression which also enables us to compute the off-diagonal elements of $\rho^{\mathrm{bath}}$, we start by rotating the embedding parameters about an angle $\theta_r$. For a fixed set of embedding parameters. This leads to the following expression:
\begin{align}
    \mathcal{F}(\theta_r) = \left(\mathcal{E}\left(u^\mathrm{T}(\theta_r)\tilde{\D},u(\theta_r)^{\dagger}\tilde{\Lambda}^cu(\theta_r)\right) \right)_{\tilde{\D}, \tilde{\Lambda}^c}
\end{align}
where $u\left(\theta_r\right) = e^{-i\theta_r\mathbf{h}_r}$ is an arbitrary unitary matrix, $\{\mathbf{h}_r\}$ is a set of hollow Hermitian matrices and the index $r \in 1,2,\dots ,\frac{B(B-1)}{2}$ runs over all non-redundant off-diagonal elements of $\Lambda^c$.
As mentioned in the main text, such a unitary transformation cannot change the energy due to the gauge invariance of the embedding Hamiltonian (EH), i.e. $\partial\F/\partial\theta_r=0$.
Exploiting this stationarity condition with respect to any unitary rotation we end up with the following equation
\begin{align} \label{ddoff}
    \frac{\partial\mathcal{F}(\theta_r)}{\partial\theta_r} \Biggr|_{\theta_r=0}=& 2\sum_{a} \left[M_r\right]_a \rho^{\mathrm{hyb}}_a 
    + \sum_{ab} \left[M_r^{c}\right]_{ab} \rho^{\mathrm{bath}}_{ba} 
    = 0
    \,.
\end{align}
The terms $\left[M_r\right]_a$ and $\left[M_r^c\right]_a$ correspond to the partial derivatives of the transformed $\tilde{\D}$ and $\tilde{\Lambda}^c$ with respect to $\theta_r$ which are given by:
\begin{align}
    \left[M_r\right]_a = \frac{\partial \left( u^\mathrm{T}(\theta_r)\tilde{\D}\right)}{\partial \theta_r} \Biggr|_{\theta_r=0} =& i\sum_b \left[\mathbf{h}_r\right]_{ba} \tilde{\D}_b  ~,\\
     \left[M_r^c\right]_a = \frac{\partial \left(u^{\dagger}(\theta_r)\tilde{\Lambda}^cu(\theta_r)\right)}{\partial \theta_r} \Biggr|_{\theta_r=0} =& i\sum_c\left[\mathbf{h}_r\right]_{ac} \left[\tilde{\Lambda}^c\right]_{cb}
     +\mathrm{H.c.}
     \,.
\end{align}
Eq. \eqref{ddoff} is exact and can be used to iteratively solve for the off-diagonal elements of $\rho^{\mathrm{hyb}}_{ab}$ since these are the only unknowns in Eq. \eqref{ddoff},

\subsection{Calculation of the Double Occupancy}

The double occupancy is not directly accessible via the gradient of $\mathcal{E}$ with respect to $\bX$. Hence, we employ a similar trick as we did for obtaining the off-diagonal elements of $\rho^{\mathrm{hyb}}_{ab}$. But instead of using the invariance of the EH with respect to unitary transformations, we employ its scaling invariance:
\begin{align}
    \E(\alpha\mathbf{X},U=\alpha) = \alpha \E(\mathbf{X},U=1)
    \,,
\end{align}
where \(\alpha\in\mathbb{R}\) is an arbitrary scaling factor.

Taking the total derivative of the above equation yields:
\begin{align}
     \frac{d \E(\alpha\mathbf{X},U=\alpha)}{d\alpha} \Biggr|_{\alpha=1} =& \E(\mathbf{X},U=1)
     \,,
\end{align}
where:
\begin{align}
    \frac{d \E(\alpha\mathbf{X},U=\alpha)}{d\alpha} = \sum_i \frac{\partial \E(\alpha\mathbf{X},U=\alpha)}{\partial X_i}X_i
    + \frac{\partial \E(\alpha\mathbf{X},U=\alpha)}{\partial U}
    \,.
\end{align}
With this, we arrive at the following expression for the double occupancy
\begin{align} \label{eq:docc}
   \frac{\partial \E(\mathbf{X},U=1)}{\partial U} = \E(\mathbf{X},U=1) -  \sum_i F_i X_i
   \,,
\end{align}
where the Hellmann-Feynman theorem has been used to evaluate the partial derivatives with respect to $X_i$ (as already described in the main text):
\begin{align}
 \frac{\partial \E(\alpha\mathbf{X},U=\alpha)}{\partial X_i} \Biggr|_{\alpha=1} &= \frac{\partial \E(\mathbf{X},U=1)}{\partial X_i} = F_i ~
 \end{align}
 and to evaluate the double occupancy as partial derivative with respect to $U$:
 \begin{align}
  \frac{\partial \E(\alpha\mathbf{X},U=\alpha)}{\partial U} \Biggr|_{\alpha=1} &= \frac{\partial \E(\mathbf{X},U=1)}{\partial U} = \bra{\Phi}c^{\dagger}_{\uparrow}c_{\uparrow}c^{\dagger}_{\downarrow}c_{\downarrow}\ket{\Phi} ~,
\end{align}
which allows us to compute the double occupancy analytically using Eq. \eqref{eq:docc}.

\section{Benchmark calculations of $\text{g}$GA+AL away from half filling}

Here we present benchmark calculations of the Hubbard model away from half-filling. Specifically, we consider dopings of 10 \%, 20 \% and 30 \%, corresponding to fillings of $N=1.1, 1.2, 1.3$. 
Similarly to the  procedure outlined in the main text, we organize our calculations as follows. 
Each series of calculations starts from a small initial value of Hubbard interaction strength $U_{\mathrm{min}}$, which is increased in steps with equal spacing $\Delta U$ up to a value $U_{\mathrm{max}}$. 
Subsequently, the interaction strength is decreased back to $U_{\mathrm{min}}$, using the same spacing. 
Below these series of calculations will be referred to as "forward sweep" and "backward sweep", respectively.

As in the main text, we will quantify the efficiency of our approach using the same:
\begin{equation}
S = \frac{N_{\text{data}}}{N_{\text{iterations}}}\,,
\end{equation}
which is the ratio of the number of times new data must be acquired and added to the database during a given calculation, \(N_{\text{data}}\), to the total number of gGA iterations necessary to perform the same calculation without ML, \(N_{\text{iterations}}\).

\subsection{Benchmarks for the Hubbard Model for the Infinite-Coordination Bethe Lattice}

\begin{figure*}
\begin{center}
    \includegraphics[width=1.0\textwidth]{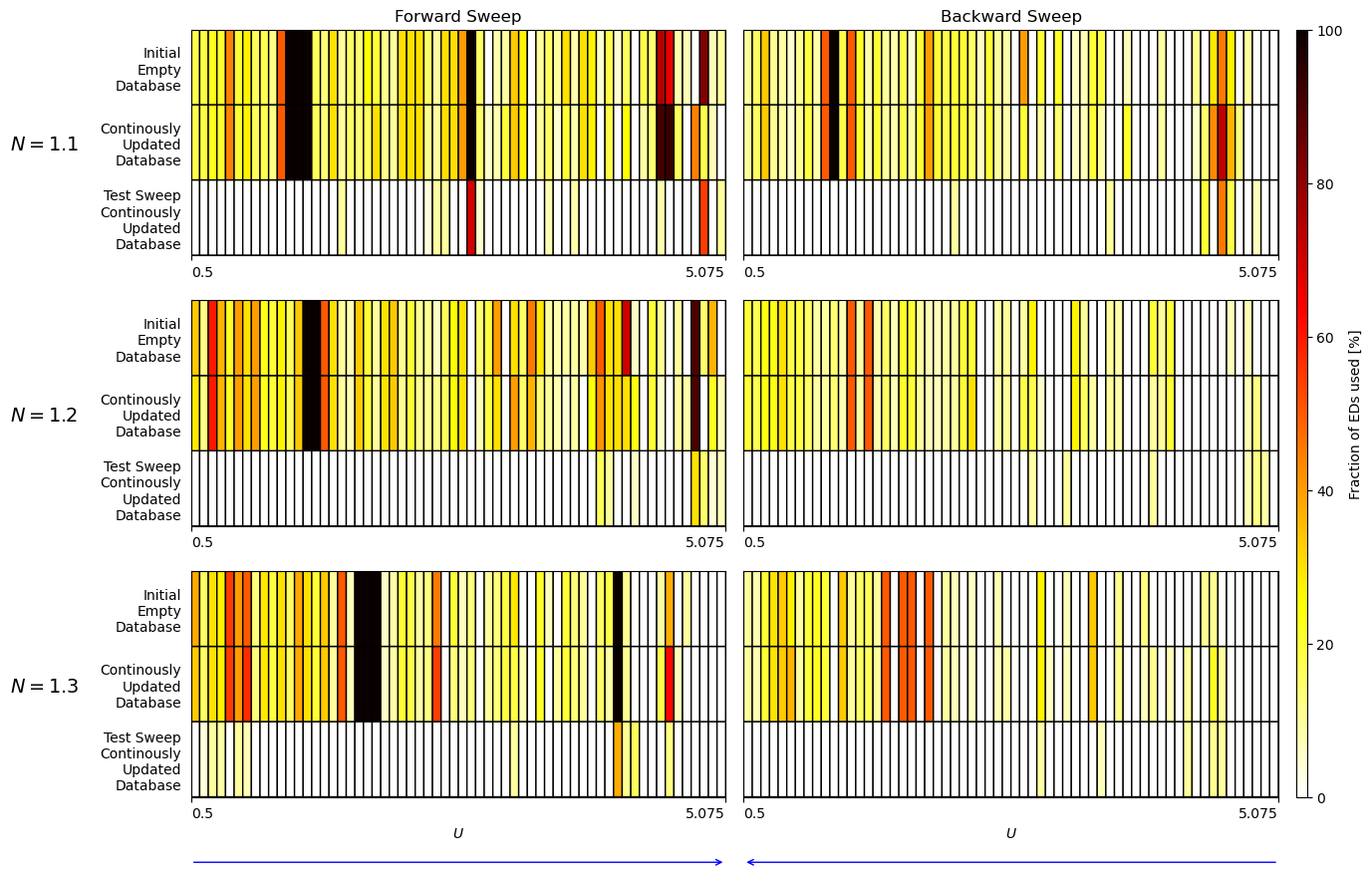}
    \caption{
Comparison of the efficiency metric \(S\) for different dopings: \(N  = 1.1, 1.2, 1.3\) for the Bethe lattice using \(\Delta U=0.075\). Each row corresponds to a different doping. The left and right columns represent the \(S\) values for forward and backward sweeps, respectively. For each \(N\), the top panel shows \(S\) calculated using an empty database at the start of each sweep, the panel in the middle displays \(S\) calculated starting from the database obtained from the half-filling calculations and updated continuously as we proceed through the series of dopings, from smallest to largest. The lower panel displays a test sweep using the database containing all of the data gathered so far.
}
    \label{fig:sc_eff}
\end{center}
\end{figure*}

In Fig. \ref{fig:sc_eff} we show the behavior of the metric $S$ for the Bethe lattice in the limit of infinite coordination number. Each row corresponds to a different number of total particles in the system and the columns show the results for the forward and backward sweep, respectively. Each row is subdivided in three part: 
The upper part refers to results obtained with an empty/reset initial training set and the middle part shows the results obtained by initializing the training database with the half-filling data and continuously updating the database throughout all of the doped calculations. 
The lower part shows the results for a second set of sweeps, to verify if the previously gathered data can be efficiently leveraged for reducing the computational cost.

We note that the efficiency metric $S$ for these calculations is even lower than for the half-filling calculations presented in the main text, indicating higher gains. The reason is that for the doped calculations the chemical potential has to be determined iteratively (see algorithm in Sec.~\ref{sec:itergGA}), rather than being fixed by particle-hole symmetry. 
Therefore, the calculation requires more ED evaluations compared to the half-filling case, which our AL algorithm can efficiently leverage on for training. 
We also note that the efficiency of the AL framework does not change substantially for the calculations performed when
continuously updating the database. 
The absence of transfer learning between calculations at different values of $N$ is not surprising, as calculations at different dopings presumably correspond to non-overlapping regions of the ambient space.

\begin{figure*}
\begin{center}
    \includegraphics[width=1.0\textwidth]{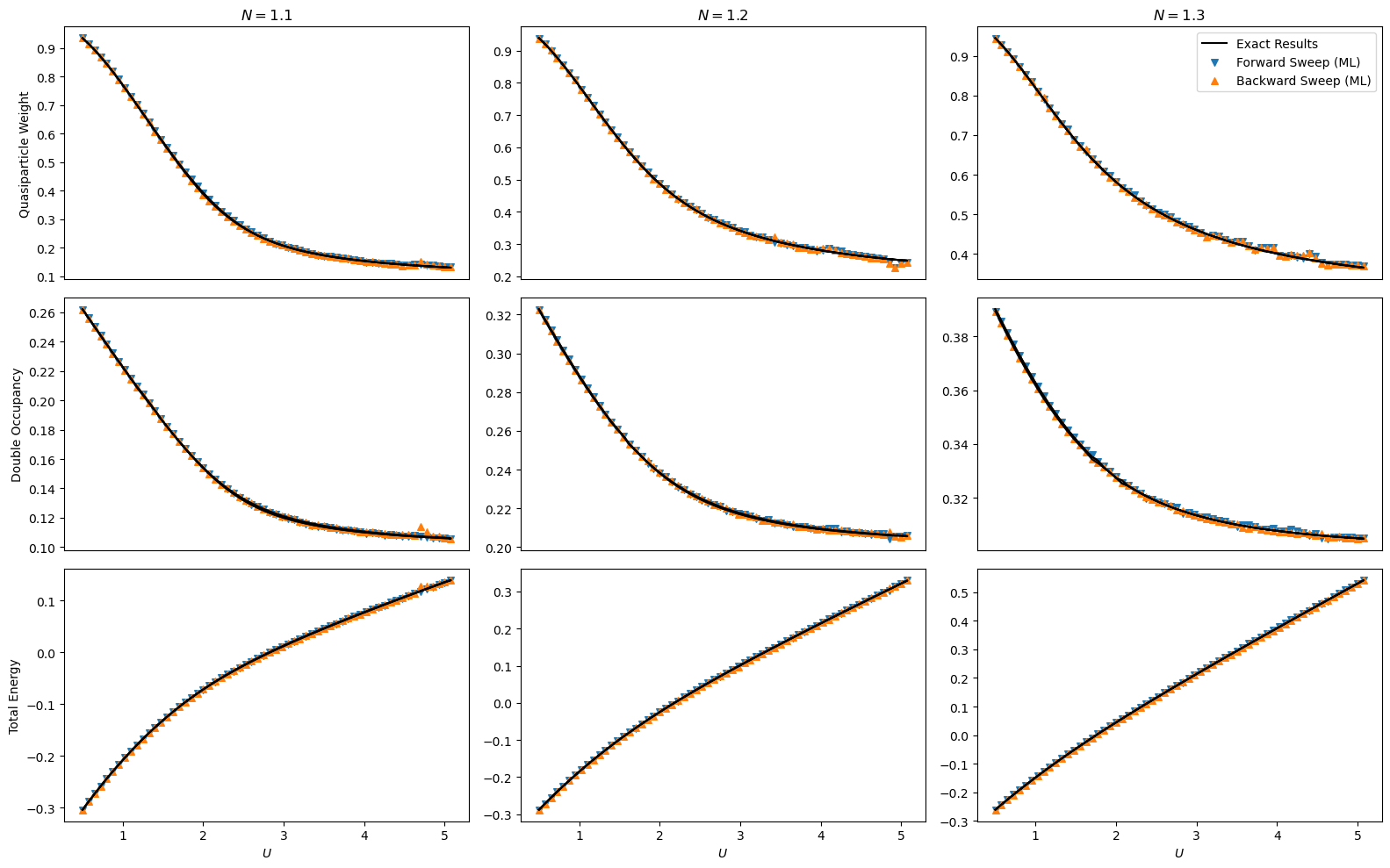}
    \caption{
    Comparison of total energy, local double occupancy, and quasi-particle weight for the Bethe lattice calculated using gGA+AL and standard gGA methods. The results referred to as "exact results" correspond to the exact gGA solutions and are represented by continuous black lines. The results obtained with gGA+AL are represented by triangles.
        }
    \label{fig:sc_acc}
\end{center}
\end{figure*}

In Fig.~\ref{fig:sc_acc} we show the predictions of the gGA+AL calculations for all fillings. The accuracy of the method is satisfactory,
although it slightly deteriorates for larger values of $U$, particularly for the quasiparticle weights $Z$.

\subsection{Benchmarks for the Hubbard Model for the 3D Cubic Lattice}

In this subsection we show calculations of the Hubbard model on a 3D cubic lattice, performing the same analysis as done above for the case of the Bethe lattice in the limit of infinite coordination.
Note that, in this section, the training database of the calculations performed with data accumulation includes all of the data gathered throughout the calculations on the Bethe lattice away from half-filling.

\begin{figure*}
\begin{center}
    \includegraphics[width=1.0\textwidth]{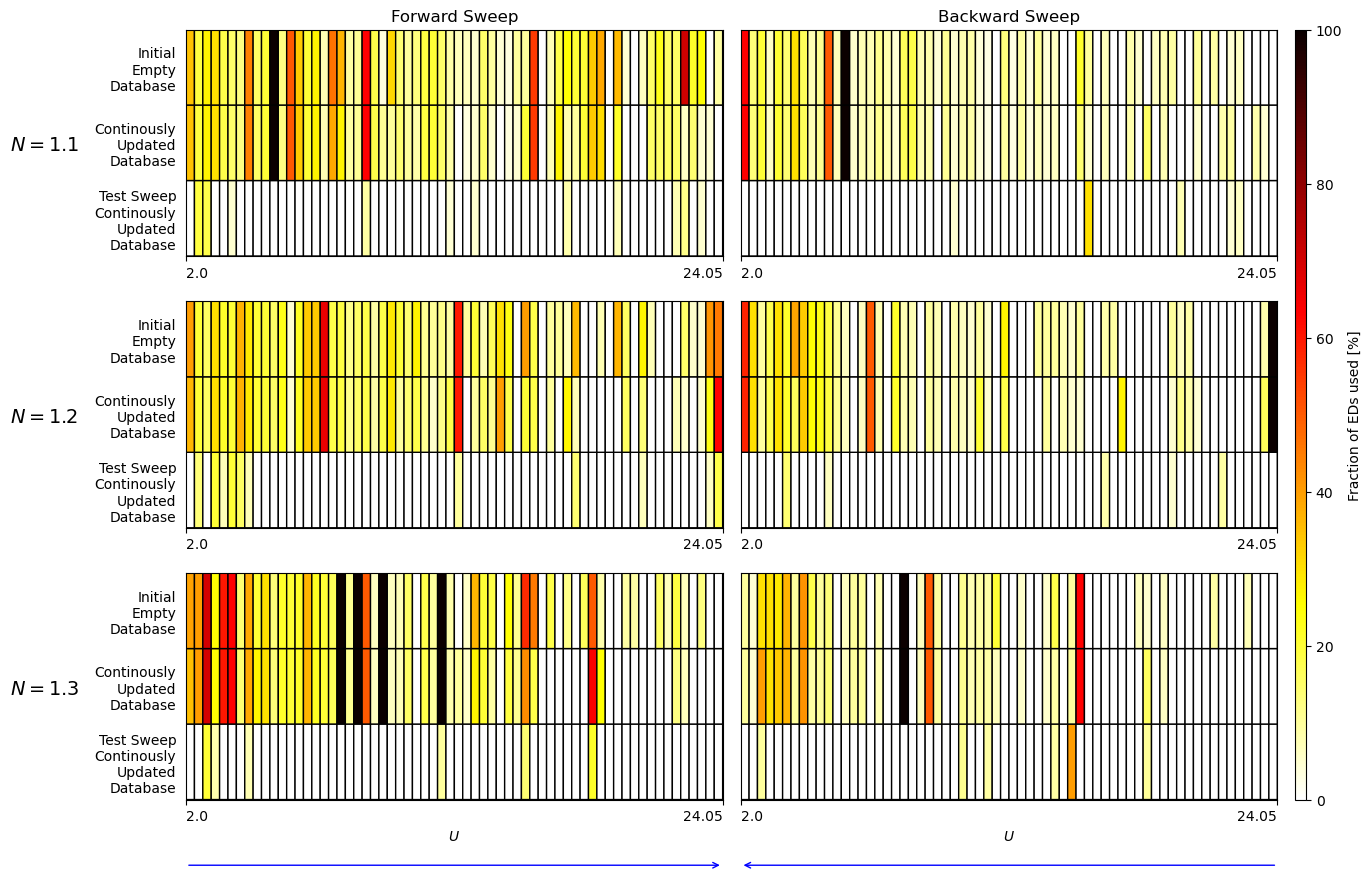}
    \caption{
    Comparison of the efficiency metric \(S\) for different dopings: \(N  = 1.1, 1.2, 1.3\) for the 3D cubic lattice using \(\Delta U=0.35\). Each row corresponds to a different doping. The left and right columns represent the \(S\) values for forward and backward sweeps, respectively. For each \(N\), the top panel shows \(S\) calculated using an empty database at the start of each sweep, the panel in the middle displays \(S\) calculated starting from the database obtained from the calculations previously performed on a Bethe-lattice and continuously updated throughout the calculations at different $N$, from smallest to largest. The lower panel displays a test sweep using the database containing all of the data gathered so far.
               }
    \label{fig:3D_eff}
\end{center}
\end{figure*}

In Fig.~\ref{fig:3D_eff} we show the behavior of the metric $S$ for the 3D cubic lattice.
As for the calculations performed for the infinite-coordination Bethe lattice, the AL method results in a substantial reduction in computational cost, even when starting each set of calculations from an empty database.
Interestingly, further gains are observed within the model of progressive data accumulation, indicating that the AL framework is able to leverage on the data previously acquired during the calculations of the model on the Bethe lattice.
We delve deeper into the underlying reason in Sec.~\ref{sec:PCA-SM}, where we analyze the data using a principal component analysis (PCA), in a similar fashion as in the main text for the half-filling calculations.

\begin{figure*}
\begin{center}
    \includegraphics[width=1.0\textwidth]{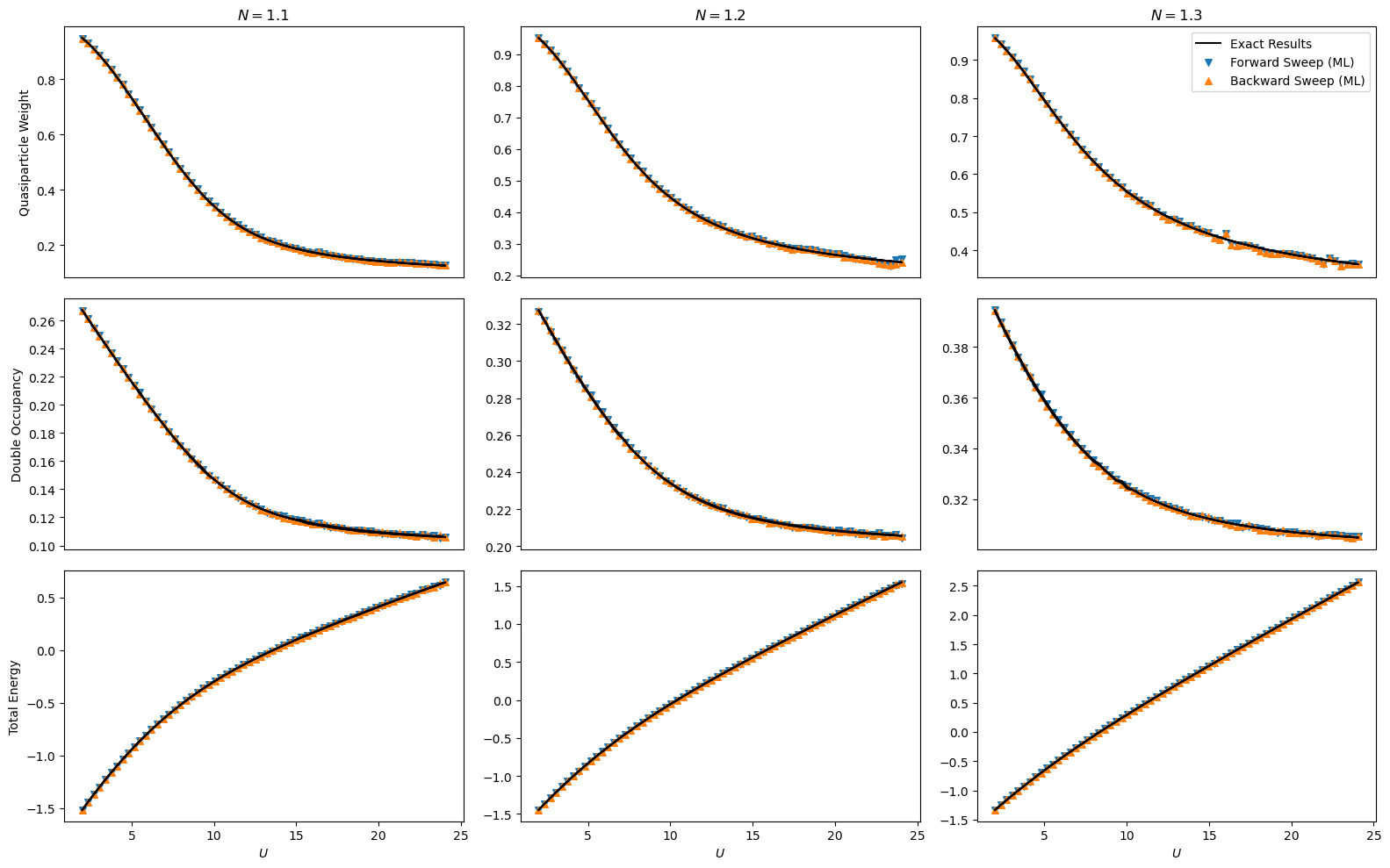}
    \caption{
    Comparison of total energy, local double occupancy, and quasi-particle weight for the 3D cubic lattice calculated using gGA+AL and standard gGA methods. The results referred to as "exact results" correspond to the exact gGA solutions and are represented by continuous black lines. The results obtained with gGA+AL are represented by triangles.
        }
    \label{fig:3D_acc}
\end{center}
\end{figure*}

In Fig.~\ref{fig:3D_acc} we show the predictions of the gGA+AL calculations for all fillings. The accuracy of the method is satisfactory, although it slightly deteriorates for larger values of $U$, as for the case of the Bethe lattice.

\subsection{Benchmarks for the Hubbard Model for the 2D Square Lattice}

Lastly, we analyze the procedure outlined above on the 2D square lattice, comparing the results obtained starting from an empty database with those obtained with the model of progressive data accumulation, including also the data acquired while performing calculations on the Hubbard model for the 3D cubic lattice.

\begin{figure*}
\begin{center}
    \includegraphics[width=1.0\textwidth]{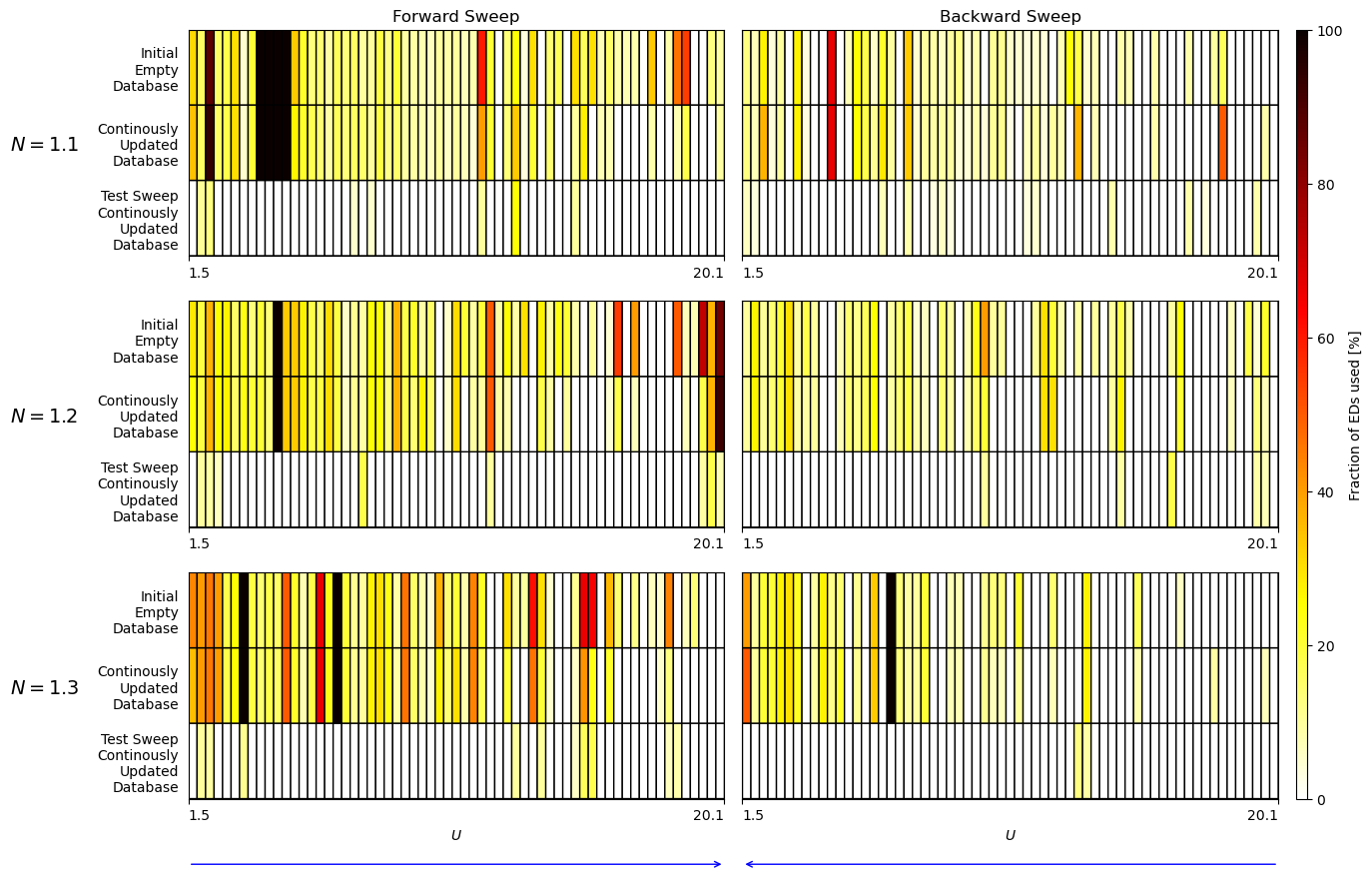}
    \caption{
        Comparison of the efficiency metric \(S\) for different dopings: \(N  = 1.1, 1.2, 1.3\) for the 2D square lattice using \(\Delta U=0.3\). Each row corresponds to a different doping. The left and right columns represent the \(S\) values for forward and backward sweeps, respectively. For each \(N\), the top panel shows \(S\) calculated using an empty database at the start of each sweep, the panel in the middle displays \(S\) calculated starting from the database obtained from the calculations previously performed on a Bethe-lattice and the 3D cubic lattice, and continuously updated throughout the calculations at different $N$, from smallest to largest. The lower panel displays a test sweep using the database containing all of the data gathered so far.
            }
    \label{fig:2D_eff}
\end{center}
\end{figure*}

In Fig.\ref{fig:2D_eff} we show the behavior of the metric $S$ for the 2D Square Lattice.
As expected, further gains are observed within the model of progressive data accumulation, indicating that the AL framework can leverage on the data previously acquired during the calculations of both the model on a Bethe lattice and the model on a 3D cubic lattice. We also note that this effect is particularly pronounced for large  values of $U$, which is the same trend observed in the main text for the gGA+AL calculations at half-filling.

\begin{figure*}
\begin{center}
    \includegraphics[width=1.0\textwidth]{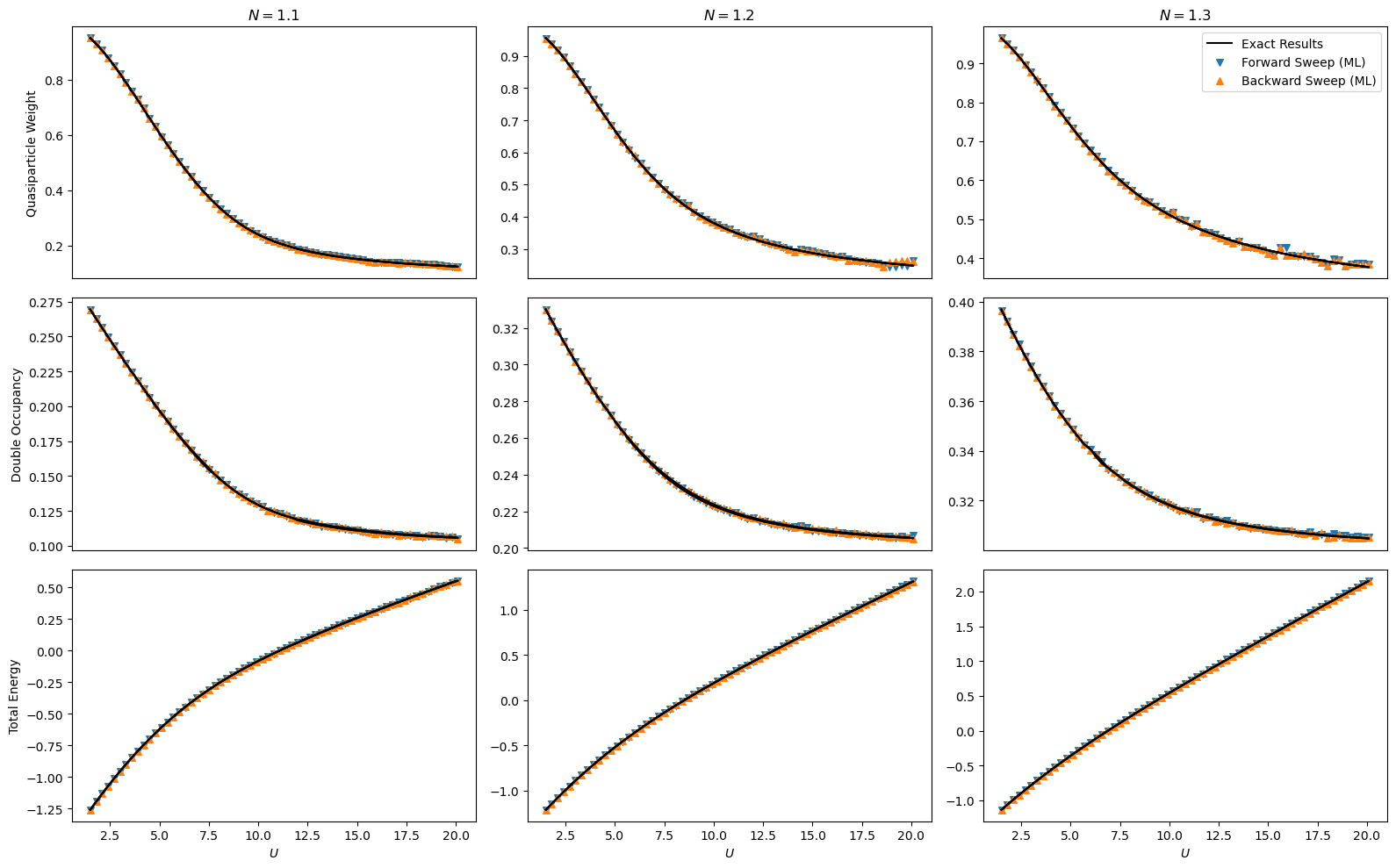}
    \caption{
    Comparison of total energy, local double occupancy, and quasi-particle weight for the 2D square lattice calculated using gGA+AL and standard gGA methods. The results referred to as "exact results" correspond to the exact gGA solutions and are represented by continuous black lines. The results obtained with gGA+AL are represented by triangles.
        }
    \label{fig:2D_acc}
\end{center}
\end{figure*}

In Fig.~\ref{fig:2D_acc} we show the predictions of the gGA+AL calculations for all fillings. The accuracy of the method is satisfactory, showing the same trends as for the other lattices analyzed previously.

\subsection{Principal Components Analysis}\label{sec:PCA-SM}

\begin{figure*}
\begin{center}
    \includegraphics[width=0.7\textwidth]{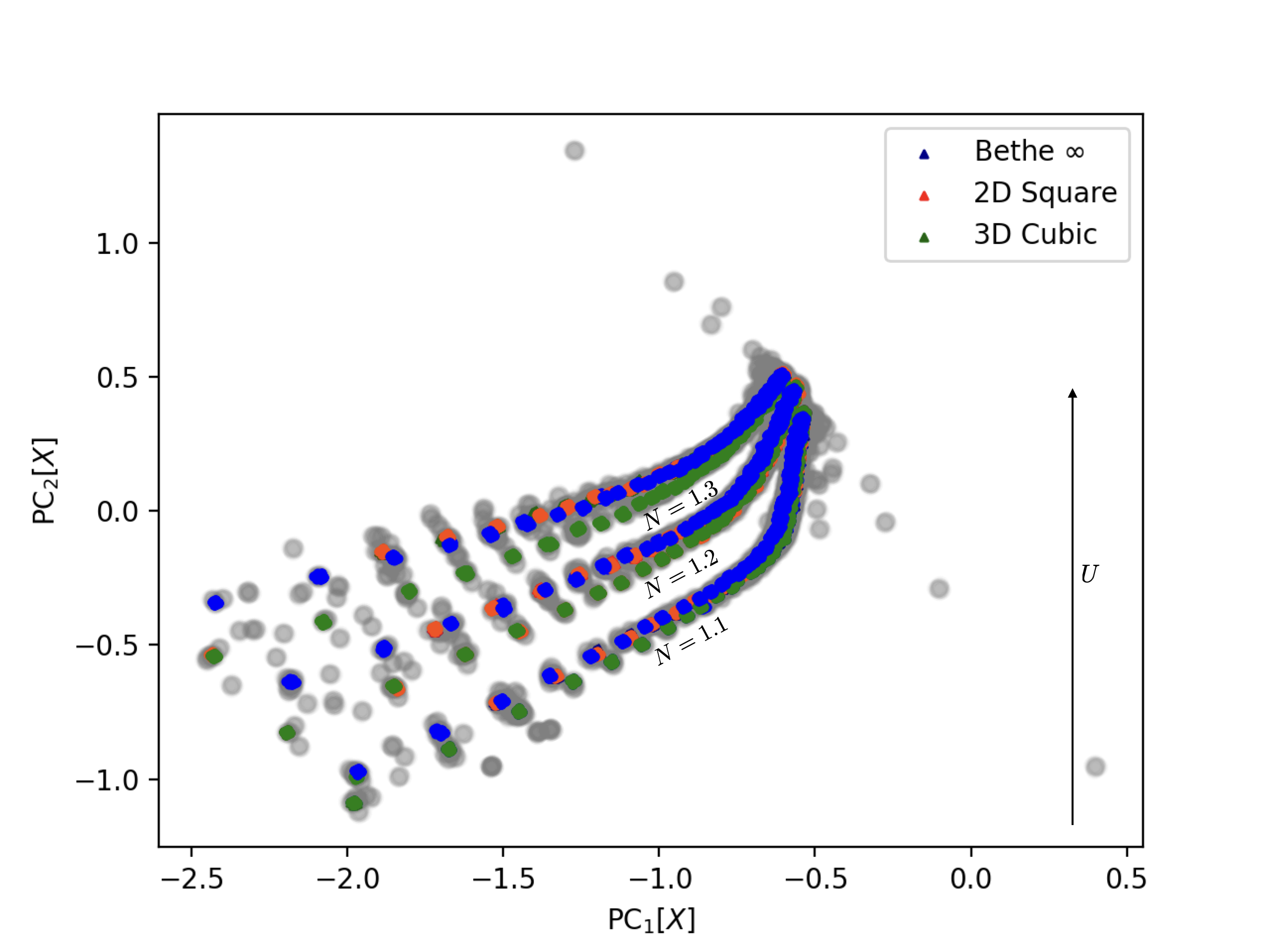}
    \caption{
    Scatter plot of the first two principal components of the training database. Points obtained after convergence for the Bethe lattice, 2D square lattice, and 3D cubic lattice are colored in blue, red, and green, respectively. All other points are colored in grey.
        }
    \label{fig:pca}
\end{center}
\end{figure*}

As in the main text, here we turn our attention to the underlying structure of the database, performing a principal component analysis of all training points accumulated during the sweeps with dopings $N=1.1, 1.2, 1.3$. 
To generate this plot each sweep  was started from an empty database.
As for the half-filling calculations discussed in the main text, we find that the first two principal components account for about 85\% of the variability of the data, which are shown in Fig.~\ref{fig:pca}.
Note that the curves obtained for different lattice structures tend to converge to the the same area at large $U$, therefore explaining the observed higher transfer-learning efficiency of our AL framework in the strongly-correlated regime.

\newpage
\clearpage


%

\end{document}